\documentclass [12pt]{article}
\usepackage {amssymb}
\usepackage {amsmath}
\usepackage [cp1251]{inputenc}
\usepackage{graphicx}
\usepackage {longtable}

\title{Large-scale instability in hydrodynamics with stable temperature stratification driven by small-scale helical force
}

\date{}

\author{ \textbf{Anatoly V.Tur},
\textbf{Vladimir V.Yanovsky}}
\begin{document}

 \maketitle
\textit{Universit\'{e} de Toulouse [UPS], CNRS, Institut de Recherche en Astrophysique et Plan\'{e}tologie,
9 avenue du Colonel Roche, BP 44346, 31028 Toulouse Cedex 4, France}

\textit{Institute for Single Crystals, Ukranian National Academy of Science, Kharkov 61001, Ukraine}

\abstract{In this work we consider the effect of a small-scale helical driving force on fluid with a stable temperature gradient with Reynolds number $Re \ll 1$. At first glance, this system   does not appear to have any instability. However, we show that large-scale vortex instability appears in fluid despite its stable stratification. In the non-linear mode, this instability gets saturated and gives a large number of stationary spiral vortex structures. Among these structures there is a stationary helical soliton and a kink of a new type. The theory is built on the rigorous asymptotical method of multi-scale development.}

\section{Introduction}

The importance of the generation processes of large-scale coherent
vortex structures in hydrodynamics is well known. When these
coherent structures appear in small-scale turbulence they play a
key role in transport processes (see for instance \cite{GEQ101}).
Numerical and laboratory experiments \cite{GEQ102}-\cite{GEQ103}
confirm the existence of coherent vortex structures, especially
for two-dimensional or quasi two-dimensional turbulence
\cite{GEQ104}-\cite{GEQ105}. Notably, they are well observed in
geophysical hydrodynamics like various cyclones in the planet's
atmospheres \cite{GEQ106}, \cite{GEQ107}. The theory of 3D
large-scale instabilities is of major interest. An example of this
theory is the generation of large-scale magnetic field by helical
small-scale turbulence ($\vec{v}rot \vec{v} \neq 0$) in the MHD
(dynamo effect or $\alpha -$effect) (see for instance
\cite{GEQ108}, \cite{GEQ109}). Many works dealt with the dynamo
theory generalization for usual hydrodynamics, and as a result it
was understood that a small-scale turbulence able to generate
large-scale perturbations cannot be simply homogeneous, isotropic
and helical \cite{GEQ110}, but must have additional special
properties. So in works \cite{GEQ111}, \cite{GEQ112} it was shown
that parity breakdown in small-scale turbulence (the external
small-scale driving forces) lead to large-scale instability, the
so-called Anisotropic Kinetic Alpha effect (AKA-effect). The
injection of the helical external force  into hydrodynamics
systems was considered in many works \cite{GEQ113}-\cite{GEQ114}.
In some cases, the existence of large-scale instability was shown
(vortex dynamo or  hydrodynamic $\alpha $-effect). So, in
particular, in work \cite{GEQ115} it is shown that the large-scale
instability exists in convective systems  with small-scale helical
turbulence. Large-scale instability was interpreted as the result
of a positive feedback loop between the poloidal and toroidal
perturbations of the velocity field, which is carried out through
the helicity coefficient. These works, as well as the results of
numerical modelling, are in details described in  review
\cite{GEQ116}, which is focused essentially on the possible
application of these results to the issue of tropical cyclone
origination.

In this work, we formulate the problem in a different way. Let us
suppose that there is a stable temperature stratification in
fluid. Let us apply to this fluid with the Reynolds number $Re \ll
1$ a small-scale, helical, external force. This force will
maintain in the fluid small-scale helical fluctuations of velocity
field ($\vec{v}rot \vec{v} \neq 0$). We consider the fluid as
boundless. At first glance, there are no instabilities at all in
this system. However, we show in this work that despite stable
stratification, a large-scale vortex instability appears in the
fluid which leads to the generation of  large-scale vortex
structures. The theory of this instability is built rigourously
using the method of asymptotical multi-scale development similar
to what was done in work of Frisch, She and Sulem for the theory
of  AKA-effect \cite{GEQ111}. In addition to the linear theory, we
also develop and study in detail the non-linear theory of this
instability saturation. We devote special attention to stationary,
non-linear, periodical vortex structures which appear as a result
of the saturation of the found instability. Among these structures
there are a spiral vortex soliton and kink of the new type.

Our work is arranged as follows: in Section 2, we set forth the
formulation of the problem and equations for stable stratification
in the Boussinesq approximation; in Section 3, we examine the
principal scheme of multi-scale development and we give secular
equations. An overall algebraic scheme of multi-scale development
over the Reynolds number up to the fifth order is described in
Appendix A. In Section 4, we find the velocity field of zero
approximation and we describe external force properties.
Calculation of terms of $R^{2} $ order is given in Section 5. The
main bulky calculations of  Reynolds stress i.e. closure of
secular equations are presented in  Appendix B. In Appendix C we
deal with some minor questions concerning the closure of
temperature equations. In Section 6 we consider the overall system
of secular equations, and we obtain and investigate equations for
the large-scale instability. In Section 7, we examine the
multi-scale development for non-linear cases. The overall
algebraic scheme of this development is given in Appendix D. In
Section 8, we calculate the velocity field of zero approximation
for the non-linear case.   The Reynolds stress calculations for
non-linear case are given in Appendix E. In Section 9, we discuss
the non-linear stage of the instability and its saturation. We
study the equations of non-linear instability and its stationary
solutions. It is shown that due to the hamiltonian nature of these
equations a large number of stationary vortex structures of the
spiral type appear. We demonstrate also that there are solutions
in the form of the spiral soliton and a kink of new type. The
obtained results are discussed in conclusions in Section 10.

\section{Main equations and formulation of the problem }

Let us consider the equations of motion of non-compressible fluid with a constant temperature gradient in  the Boussinesq approximation:

\begin{equation} \label{GEQ117} \frac{\partial \vec{V}}{\partial t} +(\vec{V}\nabla )\vec{V}=-\frac{1}{\rho _{0} } \nabla P+\nu \Delta \vec{V}+g\beta T\vec{e}_{z} +\vec{f}_{0} ; \end{equation}

\begin{equation} \label{GEQ118} \frac{\partial T}{\partial t} +(\vec{V}\nabla )T=\chi \Delta T-V_{z} A. \end{equation}
$\nabla \vec{V}=0,\vec{e}_{z} =(0,0,1)$-is the single vector in the direction of $z$ axis, $\beta $-is the thermal expansion coefficient, $A=\frac{dT_{0} }{dz} $- constant equilibrium gradient of temperature, $A=Const,A>0$. $\rho _{0} =const$. $\nabla T_{0} =A\vec{e}_{z} $- the buoyancy force  and the external force  $\vec{f}_{0} ,div\vec{f}_{0} =0$ are taken into account in Euler equation (\ref{GEQ117}).  Let us write down the force $\vec{f}_{0} $ in the form: $\vec{f}_{0} =f_{0} \vec{F}_{0} \left(\frac{x}{\lambda _{0} } ,\frac{t}{t_{0} } \right)$, where $\lambda _{0}$- characteristic scale, $t_{0} $- characteristic time, $f_{0} $- characteristic amplitude of  external force. We designate the characteristic velocity, which is engendered by external force as $v_{0} =v_{0} \left(\frac{x}{\lambda _{0} } ,\frac{t}{t_{0} } \right).$ When multiplying the first equation by the parameter  $\frac{\lambda _{0}^{2} }{\nu v_{0} } $, we choose dimensionless variables:

\[\vec{x}\to \frac{\vec{x}}{\lambda _{0} } ,t\to \frac{t}{t_{0} } ,\vec{V}\to \frac{\vec{V}}{v_{0} } ,\vec{f}_{0} \to \frac{\vec{f}_{0} }{f_{0} } ,P\to \frac{P}{\rho _{0} P_{0} } ,\]
 where
\[t_{0} =\frac{\lambda _{0}^{2} }{\nu } ,P_{0} =\frac{v_{0} \nu }{\lambda _{0} } ,f_{0} =\frac{v_{0} \nu }{\lambda _{0}^{2} } ,v_{0} =\frac{f_{0} \lambda _{0}^{2} }{\nu } .\]
In the dimensionless variables $(t,\vec{x},\vec{V})$, motion equations take the form:

\[\frac{\partial \vec{V}}{\partial t} +R(\vec{V}\nabla )\vec{V}=-\nabla P+\Delta \vec{V}+ \left( \frac{\lambda _{0}^{2} }{v_{0} \nu } \right)g\beta T\vec{e}_{z} +\vec{F}_{0} \]

\[\frac{\partial T}{\partial t} +R(\vec{V}\nabla )T=\frac{1}{\Pr } \Delta T-RV_{z} (A\lambda _{0} ),\]
where $R=\frac{\lambda _{0} v_{0} }{\nu } $ - Reynolds number on the scale $\lambda _{0} $, $\Pr =\frac{\nu }{\chi } $- is Prandtl number. We introduce the dimensionless temperature  $T\to \frac{T}{\lambda _{0} A} $ and obtain the equations system:

\[\frac{\partial \vec{V}}{\partial t} +R(\vec{V}\nabla )\vec{V}-\Delta \vec{V}=-\nabla P+\frac{Ra}{R\Pr } T\vec{e}_{z} +\vec{F}_{0} ,\]

\[\frac{1}{R} \left(\frac{\partial T}{\partial t} -\frac{1}{\Pr } \Delta T\right)=-V_{z} -(\vec{V}\nabla )T.\]
Here  $Ra=\frac{\lambda _{0}^{4} Ag\beta }{\chi \nu } $- is  Rayleigh  number on the scale $\lambda _{0} $. Further for the purpose of simplification we  will consider the case $\Pr =1$. We pass to the new temperature $T\to \frac{T}{R} $, and obtain finally:
\begin{equation} \label{GEQ119} \frac{\partial \vec{V}}{\partial t} +R(\vec{V}\nabla )\vec{V}-\Delta \vec{V}=-\nabla P+RaT\vec{e}_{z} +\vec{F}_{0} , \end{equation}

\begin{equation} \label{GEQ120} \left(\frac{\partial T}{\partial t} -\Delta T\right)=-V_{z} -R(\vec{V}\nabla )T. \end{equation}

\[div\vec{V}=0.\]
We will consider as small parameter of asymptotical development the Reynolds number $R=\frac{\lambda _{0} v_{0} }{\nu } $ ${\rm \ll }1$ on the scale  $\lambda _{0} $. The parameter $Ra$ will be considered neither big nor small, without any impact on development scheme ( i.e. outside of the scheme parameters).

Let us examine the following formulation of the problem. We consider the external force as being small and of high frequency. This force drives the small scale velocity and temperature fluctuations  on equilibrium state background. After averaging, these quickly-oscillating fluctuations equal zero. Nevertheless, the non zero terms can occur after averaging due to the fact that small non-linear interactions appear in some orders of perturbations theory. This means that they are not oscillatory, that is to say of large scale. From a formal point of view, these terms are secular, i.e. conditions for the solvability of the large scale asymptotic development. So, finding and studying the solvability equations i.e. the equations for large scales perturbations,  is actually the purpose of this work. Let us designate further small scale variables  as  $x_{0} =(\vec{x}_{0} ,t_{0} )$, and large scale ones as $X=(\vec{X},T)$. The derivative $\frac{\partial }{\partial x_{0}^{i} } $ is designated $\partial _{i} $, the derivative $\frac{\partial }{\partial t_{0} } $ is designated $\partial _{t} $,  and derivatives of large scale variables are $\frac{\partial }{\partial \vec{X}} \equiv \nabla $ i $\frac{\partial }{\partial T} \equiv \partial _{T} $ respectively.  (No misunderstanding occurs between the temperature $T$ and the large scale time $T$ since here time is argument and temperature is function).

\section{The multi-scale asymptotical development }

For constructing multi-scale asymptotic development we follow the method which is proposed in work \cite{GEQ111}. First of all,  we develop space a nd time derivatives in equations \eqref{GEQ119}, \eqref{GEQ120} into asymptotical series of the form:
\begin{equation} \label{GEQ121} \frac{\partial }{\partial x^{i} } =\partial _{i} +R^{2} \nabla +\cdots . \end{equation}

\begin{equation} \label{GEQ122} \frac{\partial }{\partial t} =\partial _{t} +R^{4} \partial _{T} +\cdots  \end{equation}
We develop the variables $\vec{V},T,P$ like so:

\begin{equation} \label{GEQ123} \vec{V}(\vec{x},t)=\vec{v}_{0} (x_{0} )+R(\vec{W}_{1} (X)+\vec{V}_{1} )+R^{2} \vec{V}_{2} +R^{3} \vec{V}_{3} +R^{4} \vec{V}_{4} +R^{5} \vec{V}_{5} +\cdots  \end{equation}
Here  $\vec{W}_{1} (X)$- is the velocity  which depends on large scale variables only.

\begin{equation} \label{GEQ124} T(\vec{x},t)=T_{0} (x_{0} )+R(\Theta _{1} (X)+T_{1} )+R^{2} T_{2} +R^{3} T_{3} +R^{4} T_{4} +R^{5} T_{5} +\cdots  \end{equation}
Here $\Theta _{1} (X)$- the temperature which depends on large scale variables only.

\begin{equation} \label{GEQ125} P(\vec{x},t)=\frac{1}{R} P_{-1} (X)+P_{0} (x_{0} )+RP_{1} +R^{2} P_{2} +R^{3} (\overline{P}_{3} (X)+P_{3} )+ \end{equation}

\[+R^{4} P_{4} +R^{5} P_{5} +\cdots \]
As we will see later, in development of pressure (\ref{GEQ125}) it is necessary to have two terms which are dependent only on large scales variables $P_{-1} (X)$ i $\overline{P}_{3} (X)$. Let us put now the developments (\ref{GEQ121})-(\ref{GEQ125}) in the equations system (\ref{GEQ119}),(\ref{GEQ120}) and write down the obtained equations up to order $R^{5} $ inclusive. The obtained equations  have a rather bulky form and are given in Appendix A. In order to simplify the writing of equations we give the algebraic structure of development only (vector indices are not written down explicitly, but can be easily restored in the necessary places). The conditions of asymptotical development solvability  (\ref{GEQ121})-(\ref{GEQ125}) of the equation system (\ref{GEQ119}), (\ref{GEQ120}) lead to the equations for the secular terms  (\ref{GEQ126}), (\ref{GEQ127}), (\ref{GEQ128}), (\ref{GEQ129}) and (\ref{GEQ130}). Let us write down the full system of secular equations:

\begin{equation} \label{GEQ131} \partial _{T} W_{1}^{k} -\Delta W_{1}^{k} +\nabla _{p} \overline{(v_{0}^{p} v_{2}^{k} +v_{0}^{k} v_{2}^{p} )}+\nabla _{p} (W_{1}^{p} W_{1}^{k} )=-\nabla _{k} \overline{P}_{3} (X); \end{equation}

\begin{equation} \label{GEQ132} \partial _{T} \Theta _{1} -\Delta \Theta _{1} =-\nabla _{p} \overline{(v_{2}^{p} T_{0} +v_{0}^{p} T_{2} )}-\nabla _{p} (W_{1}^{p} \Theta _{1} ). \end{equation}

\begin{equation} \label{GEQ133} \nabla _{p} W_{1}^{p} =0 \end{equation}

\begin{equation} \label{GEQ134} W_{1}^{z} =0 \end{equation}

\begin{equation} \label{GEQ135} \nabla P_{-1} (X)=Ra\Theta _{1} (X)\vec{l}_{z}  \end{equation}
The equations (\ref{GEQ131})-(\ref{GEQ134}) form the main equation
system. The equation (\ref{GEQ135}) is secondary and is used to
find the field $P_{-1} (X)$, which does not enter  in the main
system of equations (\ref{GEQ131})-(\ref{GEQ134}). The line above
in equations (\ref{GEQ131}), (\ref{GEQ132}) means an averaging of
corresponding terms over quick oscillations.  The calculation of
these terms (Reynolds stress) is the principal problem. After
closing of equations  system (\ref{GEQ131})-(\ref{GEQ134}) it will
describe the dynamics of large scale perturbations in this scheme.
As we can see from equations (\ref{GEQ131}), (\ref{GEQ132}), to
calculate  Reynolds stresses  we have to find fields $v_{0}^{p}
,v_{2}^{k} ,T_{0} ,T_{2} $ and to average the corresponding
products over small scale oscillations. In addition, the field
$v_{2}^{k} $ is linear depending on $\vec{W}_{1} $. That is why
the Reynolds stresses in the equation (\ref{GEQ131}) represent
expressions of the form $\beta _{kpi} \nabla _{p} W_{1}^{i} .$

Where $(i=1,2)$ and $\beta _{kpi} $- the tensor of third rank must be found. The secular condition  $W_{1}^{z} =0$ constrain the equation (\ref{GEQ131}) very strictly.

Let us write down the equation (\ref{GEQ131}) in components:

\begin{equation} \label{GEQ136} \partial _{T} W_{1}^{x} -\Delta W_{1}^{x} +\nabla _{p} (W_{1}^{p} W_{1}^{x} )+\beta _{xpx} \nabla _{p} W_{1}^{x} +\beta _{xpy} \nabla _{p} W_{1}^{y} =-\nabla _{x} \overline{P}_{3} (X); \end{equation}

\begin{equation} \label{GEQ137} \partial _{T} W_{1}^{y} -\Delta W_{1}^{y} +\nabla _{p} (W_{1}^{p} W_{1}^{y} )+\beta _{ypx} \nabla _{p} W_{1}^{x} +\beta _{ypy} \nabla _{p} W_{1}^{y} =-\nabla _{y} \overline{P}_{3} (X); \end{equation}

\begin{equation} \label{GEQ138} \beta _{zpx} \nabla _{p} W_{1}^{x} +\beta _{zpy} \nabla _{p} W_{1}^{y} =-\nabla _{z} \overline{P}_{3} (X). \end{equation}
The equation  $\nabla _{x} W_{1}^{x} +\nabla _{y} W_{1}^{y} =0$, allows us to find from the equations (\ref{GEQ136}), (\ref{GEQ137}) the pressure $\overline{P}_{3} (X)$.  But the substitution of this pressure into equation (\ref{GEQ138}) leads to a contradiction because three equations appear for two variables $W_{1}^{x} ,W_{1}^{y} $. There is just one  possibility to avoid this contradiction, which is to consider the variables $W_{1}^{x} ,W_{1}^{y} $ as functions of variable $Z$ only, i.e. $W_{1}^{x} =W_{1}^{x} (Z),W_{1}^{y} =W_{1}^{y} (Z)$. In this case, the non-linear terms in equations (\ref{GEQ136}), (\ref{GEQ137}) identically vanish, the equation $div\vec{W}_{1} =0,$ is satisfied identically and the equations  (\ref{GEQ136}),(\ref{GEQ137}) take the form:
\begin{equation} \label{GEQ139} \partial _{T} W_{1}^{x} -\Delta _{z} W_{1}^{x} +\beta _{xzx} \nabla _{z} W_{1}^{x} +\beta _{xzy} \nabla _{z} W_{1}^{y} =0 \end{equation}

\begin{equation} \label{GEQ140} \partial _{T} W_{1}^{y} -\Delta _{z} W_{1}^{y} +\beta _{yzx} \nabla _{z} W_{1}^{x} +\beta _{yzy} \nabla _{z} W_{1}^{y} =0 \end{equation}

\begin{equation} \label{GEQ141} \beta _{zzx} \nabla _{z} W_{1}^{x} +\beta _{zzy} \nabla _{z} W_{1}^{y} =-\nabla _{z} \overline{P}_{3} (X). \end{equation}
Then the velocities are determined  by the equations (\ref{GEQ139}), (\ref{GEQ140}), and the pressure is found from  the equation (\ref{GEQ141}). Taking into account the equation (\ref{GEQ132}),  the temperature $\Theta _{1} $ must also be considered as function of the variable $Z$ only.

\section{Calculations of the zero approximation fields (linear theory)}

Let us designate the operator $\partial _{t} -\partial ^{2} \equiv D_{0}$. Then, applying  this operator to the first  equation (\ref{GEQ142}), we obtain the equation only for the velocity $v_{0}$:
\begin{equation} \label{GEQ143} D_{0}^{2} v_{0}^{i} =-D_{0} \partial ^{i} P_{0} -Ra(v_{0}^{k} l^{k} )l^{i} +D_{0} F_{0}^{i}  \end{equation}
Here  $l^{i} -$vector $l^{i} =(0,0,1)$. With help of the equation $\partial _{i} v_{0}^{i} =0$, we find the pressure $P_{0} $:

\begin{equation} \label{GEQ144} P_{0} =-\frac{\partial _{p} }{\partial ^{2} } \frac{Ra}{D_{0} } l^{p} l^{k} v_{0}^{k} . \end{equation}
Eliminating the pressure  (\ref{GEQ144}) from the equation (\ref{GEQ143}), we obtain the equation for $v_{0} $:

\begin{equation} \label{GEQ145} D_{0}^{2} v_{0}^{i} =-\widehat{P}^{ip} (Ral^{p} l^{k} v_{0}^{k} )+D_{0} F_{0}^{i} , \end{equation}
Here  $\widehat{P}^{ip} -$ is the projection operator:

\begin{equation} \label{GEQ146} \widehat{P}^{ip} =\delta ^{ip} -\frac{\partial ^{i} \partial ^{p} }{\partial ^{2} }  \end{equation}
As a result, the equation (\ref{GEQ145}) can be written down in the form:
\begin{equation} \label{GEQ147} (D_{0}^{2} \delta ^{ik} +Ra\widehat{P}^{ip} l^{p} l^{k} )v_{0}^{k} =D_{0} F_{0}^{i} . \end{equation}
Let us divide this equation by $D_{0}^{2} $. Then we obtain:
\begin{equation} \label{GEQ148} (\delta ^{ik} +Ra\frac{\widehat{P}^{ip} }{D_{0}^{2} } l^{p} l^{k} )v_{0}^{k} =\frac{F_{0}^{i} }{D_{0} }  \end{equation}
Designate  operator $L_{ik} :$
\begin{equation} \label{GEQ149} L_{ik} \equiv \delta _{ik} +Ra\frac{\widehat{P}_{ip} }{D_{0}^{2} } l_{p} l_{k} . \end{equation}
Then the equation (\ref{GEQ148}) takes the form:
\begin{equation} \label{GEQ150} L_{ik} v_{0}^{k} =\frac{F_{0}^{i} }{D_{0} } , \end{equation}
And the velocity  $v_{0}^{k} $ is found using the inverse operator $L_{ik}^{-1} :$
\begin{equation} \label{GEQ151} v_{0}^{k} =L_{kj}^{-1} \frac{F_{0}^{j} }{D_{0} } . \end{equation}

\begin{equation} \label{GEQ152} L_{ik} L_{kj}^{-1} =\delta _{ij} . \end{equation}
It is easy to make sure by the direct check that the inverse operator $L_{kj}^{-1} $ has the form:

\begin{equation} \label{GEQ153} L_{kj}^{-1} =\delta _{kj} -\frac{Ra\widehat{P}_{km} l_{m} l_{j} }{D_{0}^{2} +Ra\widehat{P}_{pq} l_{p} l_{q} } . \end{equation}
Consequently the expression for the velocity  $v_{0}^{k} $ takes the form:

\begin{equation} \label{GEQ154} v_{0}^{k} =\left[\delta _{kj} -\frac{Ra\widehat{P}_{km} l_{m} l_{j} }{D_{0}^{2} +Ra\widehat{P}_{pq} l_{p} l_{q} } \right]\frac{F_{0}^{j} }{D_{0} } . \end{equation}
From the equation
\begin{equation} \label{GEQ155} D_{0} T_{0} =-l^{k} v_{0}^{k}  \end{equation}
we can find at once the field $T_{0} $:
\begin{equation} \label{GEQ156} T_{0} =-\left[1-\frac{Ra\widehat{P}_{nm} l_{m} l_{n} }{D_{0}^{2} +Ra\widehat{P}_{pq} l_{p} l_{q} } \right]\frac{(l^{j} F_{0}^{j} )}{D_{0}^{2} } . \end{equation}
In order to use these formulae we have to specify in explicit form
the helical external force $F_{0}^{j} .$The simplest and most
natural way is to specify  the external force as deterministic.
(Certainly, it is possible to specify the external force in a
statistical way with specifying random field correlator, but this
leads to more bulky calculations). As it is well known, helicity
means that $\vec{f}_{0} rot\vec{f}_{0} \ne 0$. Let us specify the
force $\vec{f}_{0} $ like so:

\begin{equation} \label{GEQ157} \vec{f}_{0} =f_{0} \left[\vec{i}\cos \varphi _{2} +\vec{j}\sin \varphi _{1} +\vec{k}(\cos \varphi _{1} +\sin \varphi _{2}) \right], \end{equation}
where

\begin{equation} \label{GEQ158} \varphi _{1} =k_{0} x-\omega _{0} t,\varphi _{2} =k_{0} y-\omega _{0} t, \end{equation}
or
\begin{equation} \label{GEQ159} \begin{array}{l} {\varphi _{1} = \vec{k}_{1} \vec{x}-\omega _{0} t, \qquad \varphi _{2} =\vec{k}_{2} \vec{x}-\omega _{0} t,} \\ {\vec{k}_{1}  = k_{0} (1,0,0);\qquad \vec{k}_{2} =k_{0} (0,1,0).} \end{array} \end{equation}
It is evident that $rot\vec{f}_{0} =k_{0} \varepsilon \vec{f}_{0} $, where $\varepsilon $-is the single pseudo scalar, i.e. helicity is equal to :

\begin{equation} \label{GEQ160} \vec{f}_{0} rot\vec{f}_{0} =k_{0} \varepsilon \vec{f}_{0}^{2} \ne 0. \end{equation}
The formulae (\ref{GEQ157}),(\ref{GEQ159}) permit us to easily
make intermediate calculations, but in the final formulae we
obviously  shall take  $f_{0} ,k_{0} ,\omega _{0} $ as equal to
one, since external force is dimensionless  and depends only on
dimensionless space and time arguments. The force (\ref{GEQ157})
is physically simple  and can be realized in laboratory
experiments and in  numerical simulation.

It is easy to write down the force (\ref{GEQ157}) in the complex form. It is evident that:
\begin{equation} \label{GEQ161} \vec{f}_{0} =\vec{A}\exp (i\varphi _{1} )+\vec{A}^{*} \exp (-i\varphi _{1} )+\vec{B}\exp (i\varphi _{2} )+\vec{B}^{*} \exp (-i\varphi _{2} ), \end{equation}
where vectors $\vec{A}$ and $\vec{B}$ has the form:

\begin{equation} \label{GEQ162} \vec{A}=\frac{f_{0} }{2} (\vec{k}-i\vec{j}),\vec{B}=\frac{f_{0} }{2} (\vec{i}-i\vec{k}), \end{equation}
and $\varphi _{1} ,\varphi _{2} $ are given by formulae (\ref{GEQ159}).

\section{Calculations of $R^{2} $ order terms }

Second approximation equations are the form (\ref{GEQ163}):
\begin{equation} \label{GEQ164} D_{0} v_{2}^{i} =-\partial ^{i} P_{2} +RaT_{2} l^{i} -\partial _{k} (W_{1}^{k} v_{0}^{i} +v_{0}^{k} W_{1}^{i} )-\partial _{k} (v_{1}^{k} v_{0}^{i} +v_{0}^{k} v_{1}^{i} ), \end{equation}

\begin{equation} \label{GEQ165} D_{0} T_{2} =-v_{2}^{k} l^{k} -\partial _{k} (W_{1}^{k} T_{0} +v_{0}^{k} \Theta _{1} )-\partial _{k} (v_{1}^{k} T_{0} +v_{0}^{k} T_{1} ) \end{equation}
Let us, as earlier, apply to the equation (\ref{GEQ164}) the operator $D_{0} $ and exclude the field $T_{2} $. For $v_{2}^{i} $ obtain:

\[D_{0}^{2} v_{2}^{i} =-D_{0} \partial ^{i} P_{2} -Ra(v_{2}^{k} l^{k} )l^{i} -\]

\[-Ral^{i} [\partial _{k} (W_{1}^{k} T_{0} +v_{0}^{k} \Theta _{1} )-\partial _{k} (v_{1}^{k} T_{0} +v_{0}^{k} T_{1} )]-\]

\[-D_{0} \partial _{k} (W_{1}^{k} v_{0}^{i} +v_{0}^{k} W_{1}^{i} )-D_{0} \partial _{k} (v_{1}^{k} v_{0}^{i} +v_{0}^{k} v_{1}^{i} ).\]
By excluding the pressure $P_{2} $ with the help of the equation $\partial _{i} v_{2}^{i} =0$, we obtain as usual the equation with the projection operator (\ref{GEQ146}):

\[D_{0}^{2} v_{2}^{i} +Ra\widehat{P}^{ip} l^{p} l^{k} v_{2}^{k} =\]

\[-\widehat{P}^{ip} \left\{Ral^{p} \partial _{k} (W_{1}^{k} T_{0} +v_{0}^{k} \Theta _{1} )+Ral^{p} \partial _{k} (v_{1}^{k} T_{0} +v_{0}^{k} T_{1} )\right\}-\]

\[-\widehat{P}^{ip} \left\{D_{0} \partial _{k} (W_{1}^{k} v_{0}^{p} +v_{0}^{k} W_{1}^{p} )+D_{0} \partial _{k} (v_{1}^{k} v_{0}^{p} +v_{0}^{k} v_{1}^{p} )\right\}.\]
We divide this equation by  $D_{0}^{2} $, and see that it can be written down in the form:

\begin{equation} \label{GEQ166} \begin{array}{l} {L_{ik} v_{2}^{k} \quad =\quad -\frac{\widehat{P}^{ip} }{D_{0}^{2} } \{ Ral^{p} \partial _{k} (W_{1}^{k} T_{0} +v_{0}^{k} \Theta _{1} )+Ral^{p} \partial _{k} (v_{1}^{k} T_{0} +v_{0}^{k} T_{1} )+} \\ {\quad \quad +D_{0} \partial _{k} (W_{1}^{k} v_{0}^{p} +v_{0}^{k} W_{1}^{p} )+D_{0} \partial _{k} (v_{1}^{k} v_{0}^{p} +v_{0}^{k} v_{1}^{p} )\} .} \end{array} \end{equation}
Where $L_{ik} $- is the former operator (\ref{GEQ149}). Taking into account the expression (\ref{GEQ153}) for the inverse operator $L_{kj}^{-1} $, we obtain:
\begin{equation} \label{GEQ167} \begin{array}{l} {v_{2}^{k} \quad =\quad -\left[\delta _{kj} -\frac{Ra\widehat{P}_{kn} l_{n} l_{j} }{D_{0}^{2} +Ra\widehat{P}_{\lambda q} l_{\lambda } l_{q} } \right]\frac{\widehat{P}^{jp} }{D_{0}^{2} } \{ Ral^{p} \partial _{m} (W_{1}^{m} T_{0} +v_{0}^{m} \Theta _{1} )+} \\ {\quad \quad +Ral^{p} \partial _{m} (v_{1}^{m} T_{0} +v_{0}^{m} T_{1} )+D_{0} \partial _{m} (W_{1}^{m} v_{0}^{p} +v_{0}^{m} W_{1}^{p} )+D_{0} \partial _{m} (v_{1}^{m} v_{0}^{p} +v_{0}^{m} v_{1}^{p} )\} .} \end{array} \end{equation}
Since $v_{1} ,T_{1,} v_{0} ,T_{0} $-are functions of only fast variables $x_{0} ,$then the second and the fourth terms in Reynolds stresses $\nabla \overline{v_{0} v_{2} }$ do not give  contribution in formula (\ref{GEQ167}) as well, we can omit them. Thereby, we will be interested in the expression:

\begin{equation} \label{GEQ168} v_{2}^{k} =-\left[\delta _{kj} -\frac{Ra\widehat{P}_{kn} l_{n} l_{j} }{D_{0}^{2} +Ra\widehat{P}_{\lambda q} l_{\lambda } l_{q} } \right]\frac{\widehat{P}^{jp} }{D_{0}^{2} } \left(Ral^{p} W_{1}^{m} \partial _{m} T_{0} +D_{0} W_{1}^{m} \partial _{m} v_{0}^{p} \right). \end{equation}
Let us write down the expression (\ref{GEQ168}) in the form:

\begin{equation} \label{GEQ169} v_{2}^{k} =-W_{1}^{m} T_{(1)}^{mk} -W_{1}^{m} T_{(2)}^{mk} , \end{equation}
Where the tensors $T_{(1)}^{mk} ,T_{(2)}^{mk} $ have the form:
\begin{equation} \label{GEQ170} T_{(1)}^{mk} =-\left[\delta _{kj} -\frac{Ra\widehat{P}_{kn} l_{n} l_{j} }{D_{0}^{2} +Ra\widehat{P}_{\lambda q} l_{\lambda } l_{q} } \right]\frac{\widehat{P}^{jp} }{D_{0}^{2} } Ral^{p} \partial _{m} T_{0} , \end{equation}
\begin{equation} \label{GEQ171} T_{(2)}^{mk} =\left[\delta _{kj} -\frac{Ra\widehat{P}_{kn} l_{n} l_{j} }{D_{0}^{2} +Ra\widehat{P}_{\lambda q} l_{\lambda } l_{q} } \right]\frac{\widehat{P}^{jp} }{D_{0} } \partial _{m} v_{0}^{p} . \end{equation}
The calculations of Reynolds stresses are rather bulky, which is why we perform them  in Appendix B.

\section{Large scale instability }

In Appendix B, we calculate Reynolds stresses  for the equation (\ref{GEQ131}); in other words, we obtain closed secular equations for large scale perturbations.  We write down these equations in the explicit form:
\begin{equation} \label{GEQ172} \partial _{T} W_{x} +\alpha \nabla _{Z} W_{y} =\nabla _{z}^{2} W_{x} , \end{equation}

\begin{equation} \label{GEQ173} \partial _{T} W_{y} -\alpha \nabla _{Z} W_{x} =\nabla _{z}^{2} W_{y}  \end{equation}

\begin{equation} \label{GEQ174} \alpha =-\varepsilon Ra\frac{4-2Ra}{(4+Ra^{2} )^{2} }  \end{equation}
Here  $\varepsilon $ designate a single pseudo scalar because expressions $\nabla _{Z} W_{y} ,-\nabla _{Z} W_{x} $ are components of $rot\vec{W}$. The equations (\ref{GEQ172}), (\ref{GEQ173}) differ from equation of AKA -effect \cite{GEQ111} by the coefficient $\alpha $ only. They obviously contain an instability which generates large scale vortex structures. Choosing the velocities $W_{x} ,W_{y} $ in the form:

\begin{equation} \label{GEQ175} W_{x} =A\exp (\gamma T)\sin kz, \end{equation}

\begin{equation} \label{GEQ176} W_{y} =B\exp (\gamma T)\cos kz, \end{equation}
we obtain the instability increment $\gamma =\pm \alpha k_{z} -k_{z}^{2} $, i.e. $\max \gamma =\frac{\alpha ^{2} }{2}$, with the $k=\frac{\alpha }{2} $. The formulae ( \ref{GEQ175}), (\ref{GEQ176}) describe spiral vortex structure (circularly polarized plane wave) with the amplitude which increases exponentially with time. These waves are sometimes called Beltrami runaways since for them there is no usual hydrodynamical interaction  $\vec{W}\nabla \vec{W}\equiv 0$.

If the external force has zero helicity, then the $\alpha $-term vanishes in accordance with the general theorem of the Reynolds stress tensor \cite{GEQ110}. Helicity is taken into account in the external force structure itself. If the temperature gradient vanishes, then it is evident that the $\alpha $-term also  vanishes. Besides the equations (\ref{GEQ172}), (\ref{GEQ173}) there is an equation to find the pressure:

\begin{equation} \label{GEQ177} C\left(W_{x} +W_{y} \right)=-\overline{P}_{3} (X)+Const, \end{equation}
where $C$
\begin{equation} \label{GEQ178} C=\frac{8-12Ra-2Ra^{2} -3Ra^{3} -Ra^{4} }{(4+Ra^{2} )^{3} } , \end{equation}
and also the equation for the large scale temperature perturbation
found in Appendix C:

\begin{equation} \label{GEQ179} \partial _{T} \Theta _{1} -\Delta \Theta _{1} =-2\frac{Ra-2}{(4+Ra^{2} )^{2} } \nabla _{Z} \left(W_{x} +W_{y} \right) \end{equation}
and the equation to find the pressure  $P_{-1} (X):$

\begin{equation} \label{GEQ180} \nabla _{z} P_{-1} (X)=Ra\Theta _{1} (X) \end{equation}
It is easy to solve all these equations with the known fields $W_{x} ,W_{y} .$

It should be noted that the field $W_{1}^{z} $ is equal to zero in the main approximation only. $W^{z} \ne 0$ in the approximation of higher orders. In compliance with this  in the approximation of higher orders appear also derivatives $\frac{\partial }{\partial X} ,\frac{\partial }{\partial Y} $. This means, that in fact one has to consider that

\begin{equation} \label{GEQ181} \frac{\partial }{\partial Z} {\rm \gg }\frac{\partial }{\partial X} ,\frac{\partial }{\partial Y} . \end{equation}
The conditions  (\ref{GEQ181}) means that the horizontal scales of the instable vortices  $L_{x} ,L_{y} {\rm \gg }L_{z} $ (vertical scale). $L_{z} $ ${\rm \gg }\lambda _{0} $.

Since the increasing  fields $W_{x} ,W_{y} $ belong to Beltrami type,  the general non linearity ($W\nabla )W$cannot limit their increase, because, as it has been already stated, it is identically equal to zero. In order to describe the process of this  instability saturation it is necessary to develop a non linear theory of  $\alpha $-effect. This theory is described in the following  sections.

\section{Multi scale development for the non linear case }

In the non linear case the large scale field $W(X)$ is no small any longer that is why the asymptotical development (\ref{GEQ123})-(\ref{GEQ125}) has to be modified. Let us search the solution for the equations (\ref{GEQ119}), (\ref{GEQ120}) in the following form:

\begin{equation} \label{GEQ182} \vec{V}(\vec{x},t)=\frac{1}{R} \vec{W}_{-1} (X)+\vec{v}_{0} (x_{0} )+R\vec{V}_{1} +R^{2} \vec{V}_{2} +R^{3} \vec{V}_{3} +\cdots  \end{equation}

\begin{equation} \label{GEQ183} T(\vec{x},t)=\frac{1}{R} T_{-1} (X)+T_{0} (x_{0} )+RT_{1} +R^{2} T_{2} +R^{3} T_{3} +\cdots  \end{equation}

\begin{equation} \label{GEQ184} P(\vec{x},t)=\frac{1}{R^{3} } P_{-3} (X)+\frac{1}{R^{2} } P_{-2} (X)+\frac{1}{R} P_{-1} (X)+P_{0} (x_{0} )+R(P_{1} +\overline{P}_{1} (X))+R^{2} P_{2} +R^{3} P_{3} +\cdots  \end{equation}
For the derivatives we use the previous development (\ref{GEQ121}),(\ref{GEQ122}). Substituting these expressions into the initial equations (\ref{GEQ119}),(\ref{GEQ120}) and gathering together the terms of the same order $R$, up to degree $R^{3} $ inclusive, we obtain the equations of multi scale asymptotical development. The algebraic scheme of this development is given in Appendix D. The result of this scheme is the separation of secular terms from fast oscillating ones. Let us give secular terms in the explicit form:

\begin{equation} \label{GEQ185} \partial _{T} W^{i} -\Delta W^{i} +\nabla _{k} \left(\overline{v_{0}^{k} v_{0}^{i} }\right)=-\nabla _{i} \overline{P}_{1} ; \end{equation}

\begin{equation} \label{GEQ186} \partial _{T} T-\Delta T+\nabla _{k} (\overline{v_{0}^{k} T_{0} })=0. \end{equation}
In these equations we do not write the law index $(-1)$. Besides this, there are secular equations:

\begin{equation} \label{GEQ187} \nabla _{i} W^{i} =0,W^{z} =0, \end{equation}

\begin{equation} \label{GEQ188} \nabla _{k} (W^{k} W^{i} )=-\nabla _{i} P_{-1} , \end{equation}

\begin{equation} \label{GEQ189} \nabla _{k} (W^{k} T)=0. \end{equation}
The equations  (\ref{GEQ187})-(\ref{GEQ189}) are satisfied in the previous geometry:

\begin{equation} \label{GEQ190} W=(W^{x} (z),W^{y} (z),0),and,P_{-1} =Const. \end{equation}

There is also an equation to find the pressure $P_{-3} :$

\begin{equation} \label{GEQ191} \nabla _{z} P_{-3} =RaTl_{z} . \end{equation}
It is clear that  the essential equations for finding the non linear alpha-effect is the equation (\ref{GEQ185}). In order to obtain these equations in the closed form  we need to calculate the  Reynolds  stresses $\nabla _{k} \left(\overline{v_{0}^{k} v_{0}^{i} }\right)$. Below, we will deal with the solution of this problem. First, we have to calculate the $v_{0}^{k} $ fields of zero approximation. From the asymptotical development in zero order we have the equations:

\begin{equation} \label{GEQ192} \partial _{t} v_{0}^{i} -\partial ^{2} v_{0}^{i} +W^{k} \partial _{k} v_{0}^{i} =-\partial _{i} P_{0} +RaT_{0} l^{i} +F_{0}^{i} , \end{equation}

\begin{equation} \label{GEQ193} \partial _{t} T_{0} -\partial ^{2} T_{0} +W^{k} \partial _{k} T_{0} =-v_{0}^{k} l^{k} . \end{equation}

\section{Calculation of zero approximation fields in the non linear case }

Let us introduce the operator $\widehat{D}_{0} :$

\begin{equation} \label{GEQ194} \widehat{D}_{0} =\partial _{t} -\partial ^{2} +W^{k} \partial _{k} . \end{equation}
Using the operator $\widehat{D}_{0} $, we write down the equations (\ref{GEQ192}) and (\ref{GEQ193}) in the form:

\begin{equation} \label{GEQ195} \widehat{D}_{0} v_{0}^{i} =-\partial _{i} P_{0} +RaT_{0} l^{i} +F_{0}^{i} , \end{equation}

\begin{equation} \label{GEQ196} \widehat{D}_{0} T_{0} =-v_{0}^{k} l^{k}  \end{equation}
Eliminating the temperature from the equation (\ref{GEQ195}) we obtain:

\begin{equation} \label{GEQ197} \widehat{D}_{0}^{2} v_{0}^{i} =-\widehat{D}_{0} \partial _{i} P_{0} -Ra(v_{0}^{k} l^{k} )l^{i} +\widehat{D}_{0} F_{0}^{i} . \end{equation}

\begin{equation} \label{GEQ198} \partial _{i} v_{0}^{i} =0. \end{equation}
Eliminating the pressure from the equation (\ref{GEQ197})we obtain:

\begin{equation} \label{GEQ199} \widehat{D}_{0}^{2} v_{0}^{i} =-\widehat{P}_{ip} (Rav_{0}^{k} l^{k} l^{p} )+\widehat{D}_{0} F_{0}^{i}  \end{equation}

or:
\begin{equation} \label{GEQ200} (\widehat{D}_{0}^{2} \delta _{ik} +\widehat{P}_{ip} Ral^{k} l^{p} )v_{0}^{k} =\widehat{D}_{0} F_{0}^{i} . \end{equation}
Dividing this equation by $\widehat{D}_{0}^{2} $, we can write it in the form:

\begin{equation} \label{GEQ201} L_{ik} v_{0}^{k} =\frac{F_{0}^{i} }{\widehat{D}_{0} } , \end{equation}
Where $L_{ik} $ is the prior operator (\ref{GEQ149}) in which $D_{0} $ is replaced by $\widehat{D}_{0} $. In much the same way the inverse operator also coincides with (\ref{GEQ153}) when substitung $D_{0} $ by $\widehat{D}_{0} $. As a result it is easy to find the fields $v_{0}^{k} $ and $T_{0} $:
\begin{equation} \label{GEQ202} v_{0}^{k} =\left[\delta _{kj} -\frac{Ra\widehat{P}_{km} l_{m} l_{j} }{\widehat{D}_{0}^{2} +Ra\widehat{P}_{pq} l_{p} l_{q} } \right]\frac{F_{0}^{j} }{\widehat{D}_{0} } , \end{equation}

\begin{equation} \label{GEQ203} T_{0} =-\left[1-\frac{Ra\widehat{P}_{nm} l_{m} l_{n} }{\widehat{D}_{0}^{2} +Ra\widehat{P}_{pq} l_{p} l_{q} } \right]\frac{(l^{j} F_{0}^{j} )}{\widehat{D}_{0}^{2} } . \end{equation}
The external force $\vec{F}_{0} $ has the prior form (\ref{GEQ161}). The effect of the operator $\widehat{D}_{0} $ on proper function $\exp (i\omega t+i\vec{k}\vec{x})$ has obviously the form:

$\widehat{D}_{0} \exp (i\omega t+i\vec{k}\vec{x})=\widehat{D}_{0} (\omega ,\vec{k})\exp (i\omega t+i\vec{k}\vec{x})$, where $\widehat{D}_{0} (\omega ,\vec{k})$ is:

\begin{equation} \label{GEQ204} \widehat{D}_{0} (\omega ,\vec{k})=i(\omega +\vec{k}\vec{W})+k^{2} . \end{equation}
From this it is evident that:
\begin{equation} \label{GEQ205} \widehat{D}_{0} (\omega ,-\vec{k}_{1} )=i(\omega -\vec{k}_{1} \vec{W})+k_{1}^{2} , \end{equation}
\begin{equation} \label{GEQ206} \widehat{D}_{0}^{*} (\omega ,-\vec{k}_{1} )=\widehat{D}_{0} (-\omega ,\vec{k}_{1} ), \end{equation}
\begin{equation} \label{GEQ207} \widehat{D}_{0} (\omega ,-\vec{k}_{2} )=i(\omega -\vec{k}_{2} \vec{W})+k_{2}^{2} , \end{equation}
\begin{equation} \label{GEQ208} \widehat{D}_{0}^{*} (\omega ,-\vec{k}_{2} )=\widehat{D}_{0} (-\omega ,\vec{k}_{2} ). \end{equation}
From the formulae (\ref{GEQ202}) and (\ref{GEQ161}) follows that the field $v_{0}^{k} $ is composed of four terms:$v_{0}^{k} =v_{01}^{k} +v_{02}^{k} +v_{03}^{k} +v_{04}^{k} $ where

\[v_{02}^{k} =(v_{01}^{k} )^{*} ,v_{04}^{k} =\left(v_{03}^{k} \right)^{*} ,\]
\begin{equation} \label{GEQ209} v_{01}^{k} =e^{i\varphi _{1} } \left[\delta _{kj} -\frac{Ra\widehat{P}_{km} l_{m} l_{j} }{\widehat{D}_{0}^{2} (-\omega _{0} ,\vec{k}_{1} )+Ra\widehat{P}ll} \right]\frac{A^{j} }{\widehat{D}_{0} (-\omega _{0} ,\vec{k}_{1} )} , \end{equation}
\begin{equation} \label{GEQ210} v_{03}^{k} =e^{i\varphi _{2} } \left[\delta _{kj} -\frac{Ra\widehat{P}_{km} l_{m} l_{j} }{\widehat{D}_{0}^{2} (-\omega _{0} ,\vec{k}_{2} )+Ra\widehat{P}ll} \right]\frac{B^{j} }{\widehat{D}_{0} (-\omega _{0} ,\vec{k}_{2} )} . \end{equation}
As it was said earlier, in scalar operators $\widehat{D}_{0} $ one can take $\omega _{0} =1,\vec{k}_{1} =(1,0,0),\vec{k}_{2} =(0,1,0)$. Then taking into account that $\widehat{P}ll=1$, we obtain:

\begin{equation} \label{GEQ211} \widehat{D}_{0} (\omega _{0} ,\vec{-}k_{1} )=1+i(1-W_{1} )\equiv D_{1} , \end{equation}

\[\widehat{D}_{0} (-\omega _{0} ,\vec{k}_{1} )=D_{1}^{*} \]

\begin{equation} \label{GEQ212} \widehat{D}_{0} (\omega _{0} ,-\vec{k}_{2} )=1+i\left(1-W_{2} \right)\equiv D_{2} , \end{equation}
\[\widehat{D}_{0} (-\omega _{0} ,\vec{k}_{2} )=D_{2}^{*} ..\]
Taking into consideration these formulae we can write down the velocities  $v_{0}^{k} $ in the form:

\begin{equation} \label{GEQ213} v_{01}^{k} =e^{i\varphi _{1} } \left[\delta _{kj} -\frac{Ra\widehat{P}_{km} l_{m} l_{j} }{D_{1}^{*2} +Ra} \right]\frac{A^{j} }{D_{1}^{*} } , \end{equation}

\begin{equation} \label{GEQ214} v_{03}^{k} =e^{i\varphi _{2} } \left[\delta _{kj} -\frac{Ra\widehat{P}_{km} l_{m} l_{j} }{D_{2}^{*2} +Ra} \right]\frac{B^{j} }{D_{2}^{*} } . \end{equation}

\section{Non linear instability and large scale vortex structures}

The calculations of the Reynolds stresses for the non linear case are performed in the  Appendix E. Let us write down in the explicit form the equations for the non linear instability:

\begin{equation} \label{GEQ215} \partial _{T} W_{1} -\nabla _{z}^{2} W_{1} =-\nabla _{z} T_{(2)}^{31} = \end{equation}

\[=\nabla _{z} \frac{Ra(1-W_{2} )}{[1+(1-W_{2} )^{2} ][\left(W_{2} (2-W_{2} )+Ra\right)^{2} +4(1-W_{2} )^{2} ]} ,\]

\begin{equation} \label{GEQ216} \partial _{T} W_{2} -\nabla _{z}^{2} W_{2} =-\nabla _{z} T_{(1)}^{32} = \end{equation}
\[=-\nabla _{z} \frac{Ra(1-W_{1} )}{[1+(1-W_{1} )^{2} ][\left(W_{1} (2-W_{1} )+Ra\right)^{2} +4(1-W_{1} )^{2} ]} .\]
Here we introduced the following notations: $W_{1} \equiv W_{x} ,W_{2} \equiv W_{y} .$ It is easy to verify that with small values of the variables  $W_{1} ,W_{2} $ the equations  (\ref{GEQ215}), (\ref{GEQ216}) are reduced to the equations (\ref{GEQ172})-(\ref{GEQ174}) and describe the linear stage of instability. It is clear that with the increasing of $W_{1} ,W_{2} $ the non linear terms decrease and the instability gets saturated. As a result of the development and stabilization of instability the non linear vortex helical structures appear. The study of the form of these stationary structures is of interest. For that purpose we take $\partial _{T} W_{1} =\partial _{T} W_{2} =0$ in the equations  (\ref{GEQ215}), (\ref{GEQ216}). Integrating these equation over $z$, we obtain:

\begin{equation} \label{GEQ217} \frac{\partial X}{\partial z} =\frac{RaP}{\left(1+P^{2} \right)\left[4P^{2} +(1-P^{2} +Ra)^{2} \right]} +C_{1} , \end{equation}

\begin{equation} \label{GEQ218} \frac{\partial P}{\partial z} =-\frac{RaX}{\left(1+X^{2} \right)\left[4X^{2} +(1-X^{2} +Ra)^{2} \right]} -C_{2} . \end{equation}
Here new variables are introduced  $X=1-W_{1} ,P=1-W_{2} $, and $C_{1} ,C_{2} $- are integration constants. The system of equations (\ref{GEQ217}) and (\ref{GEQ218}) can be write down in the hamiltonian form:

\begin{equation} \label{GEQ219} \frac{\partial X}{\partial z} =\frac{\partial H}{\partial P} ,\frac{\partial P}{\partial z} =-\frac{\partial H}{\partial X} . \end{equation}
Here the variable $z$ plays the role of time and the hamiltonian $H$ has the form:

\begin{equation} \label{GEQ220} H=U(P)+U(X)+C_{1} P+C_{2} X+C_{3} . \end{equation}
Where $U(x)$ has the form:

\begin{equation} \label{GEQ221} U(x)=\frac{1}{4(4+Ra)} \ln \frac{(1+x^{2} )^{2} }{4Ra+(x^{2} +1-Ra)^{2} } +\frac{\sqrt{Ra} }{4(4+Ra)} \arctan \frac{1+x^{2} -Ra}{2\sqrt{Ra} } . \end{equation}
The function $H$ (\ref{GEQ220}),(\ref{GEQ221}) is obviously the first integral of the equations system (\ref{GEQ217}),(\ref{GEQ218}) and can be found by the direct integration of this system. With $C_{1} =0,C_{2} =0$ the function $U(x)$ is limited above and below as well. That is why the hamiltonian section by the constant $H=H_{0} ,$ gives closed periodical trajectories on the phase plane  $(X,P)$, which correspond to the helical vortex structures in the real space.
\begin{figure}
   \centerline{\includegraphics[width=8 cm]{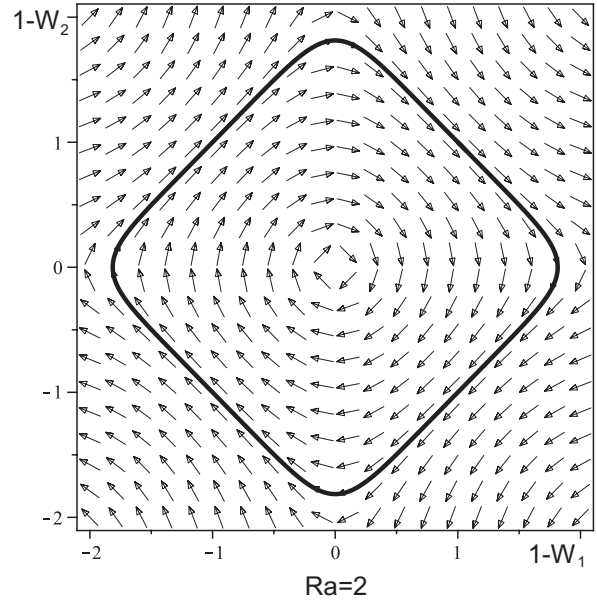}}
  \caption{Phase picture of the dynamical system with $Ra=2$, $C_1=C_2=0$. The bold line shows the phase trajectory which comes out of the point $(1, 1)$ and after the "time" $Z=L$ comes back to the same point. This trajectory presents the stationary solution of the boundary problem with the rigid boundaries in the layer whose thickness is $L=z$.}\label{fig1}
\end{figure}
Examples of phase pictures for $Ra=2$ and $Ra=3$ are represented in fig.1 and fig.2.  With $C_{1} =0,C_{2} =0$ on the phase plane there is only one elliptical point. Closed trajectories correspond to the periodical non linear vortex structures. Thick closed lines correspond to the non linear structures which are also the solutions of  the boundary problem with the rigid boundary :

\[W_{1} =0,W_{2} =0,z=0,z=L,\]
where $L$ is the period over $z$ of phase trajectory, which gets out with  $z=0$, of  $W_{1} =0,W_{2} =0$ and gets back to the same point with $z=L$. The space structures of periodical solutions is presented in  fig.3-fig.5. If one of the constants, for instance $C_{1} \ne 0$, then one hyperbolic point appears on phase picture. For instance, phase pictures with $C_{1} =0.1$ are presented in fig.6. The example of periodical vortex structure which corrersponds to the closed trajectory on pase plane with  $Ra=2$ is given in fig.7.  The solution which corresponds to the separatrix in fig.6 is of particular interest. This solution describes the solitary spiral turn of the velocity field around the axis $z$ (soliton) see fig 8. Moving away from soliton the velocity field becomes constant. This kind of solitons were not  known earlier.  The interesting particularity of this soliton is the fact that it is also the solution of the boundary problem with free boundaries. For this boundary problem \cite{GEQ222}:
\begin{figure}
  \centerline{\includegraphics[width=8 cm]{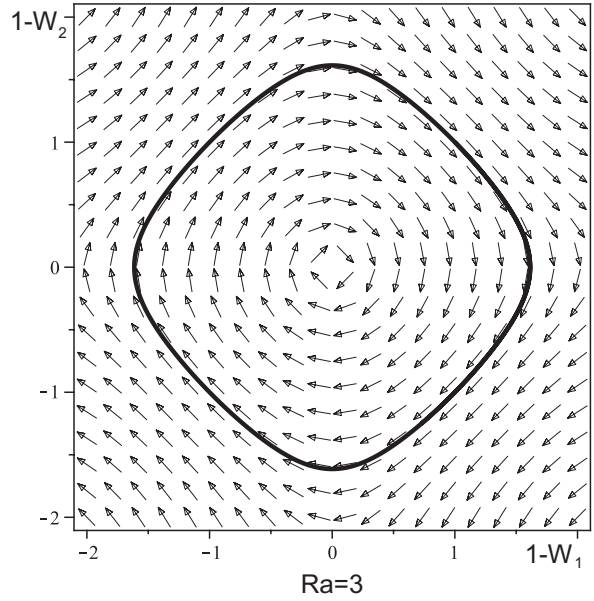}}
  \caption{Phase picture of the dynamical system with $Ra=3$, $C_1=0$, $C_2=0$. The bold line shows the trajectory which corresponds to the stationary solution of the boundary problem with rigid boundaries with $z=0$ and $z=L$.}\label{fig2}
\end{figure}

\[\frac{\partial W_{1} }{\partial z} =\frac{\partial W_{2} }{\partial z} =0\]
on the fluid boundary. In addition to that, boundaries must be on
a big distance from the soliton, much bigger than the soliton's
characteristic dimensions. In cases when there are two constants
$C_{1} \ne 0,C_{2} \ne 0$  two hyperbolic and two elliptical
points appear on phase picture. The example of this phase picture
with $C_{1} =0.1,C_{2} =0.1$ is shown in fig.9. As earlier, the
periodic vortex structures correspond to closed trajectories
around elliptical points. Localized solutions (solitons)
correspond to a separatrix in fig.9. Since the separatrix connects
two different hyperbolic points, the soliton now has two different
limiting values, with $z\to \pm \infty $ ,fig.10. This soliton is
called a kink. Thereby, spiral kinks correspond to the separatrix
in  fig.9. These kinks are also solutions of the boundary problem
with free boundaries. In conclusion, it should be remembered that
the system of the equations (\ref{GEQ215}),(\ref{GEQ216}) is
closed. The velocity field $W_{1} ,W_{2} $ determines the pressure
$\overline{P}_{1} $, according to the formulae:

\[-\overline{P}_{1} =T_{(1)}^{33} +T_{(2)}^{33} ,\]
where $T_{(1)}^{33} ,T_{(2)}^{33} $ are given by the formulae (\ref{GEQ354}),(\ref{GEQ355}). Besides, the velocity field $W_{1} ,W_{2} $ gives the contribution to the equation for temperature  (\ref{GEQ186}). Closure of this equation is made in much the same way as the closure for velocity. Nevertheless, this equation is secondary and here we do not give the result of this closure.

\section{Conclusions and discussion of  results}

In this work, it is shown that in fluid with stable
stratification, a large scale instability appears under the action
of small scale helical force. The result of instability is the
generation of vortex structures of the Beltrami type. The vortices
have the characteristic vertical dimension $L_{z} {\rm \gg
}\lambda _{0} $ and the horizontal dimension is much bigger than
the vertical one. 
\begin{figure}
  \centerline{\includegraphics[width=7 cm]{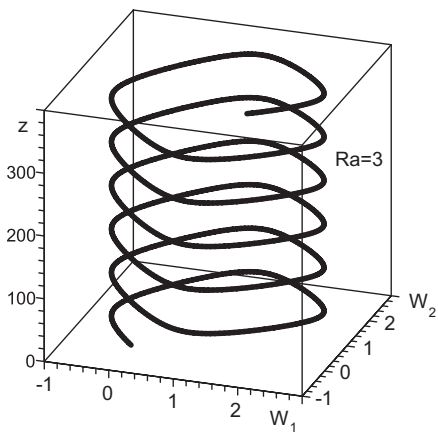}}
  \caption{Helical vortex structure with $Ra=3$, $C_1=0$, $C_2=0$.}\label{fig3}
\end{figure}
Since the vertical component of the velocity
$W_{z} $ is equal to zero in the main approximation and the
stratification is stable, then  the found instability has no
relation with convection. The structure of the equation which
describes the instability in linear approximation is the same as
the equation of $\alpha $-effect or more precisely as the equation
of AKA-effect. As a result, instability generates plane spiral
waves with circular polarisation (Beltrami runaway). With an
increase in amplitude, the instability and its stabilization are
described by a non linear theory. Stationary equations appear to
be hamiltonian, which is why they are a rich set of periodical
spiral vortex structures. Notwithstanding the fact that attention
in this work was essentialy paid to a boundary free problem,  it
should be noted that some periodical solutions  turn out to be
solutions of the boundary problem with rigid boundaries. We would
like to pay special attention to stationary soliton and kink,
which correspond to the separatrix on the phase plane. This is the
solitons of the new type. In real space it describes one spiral
turn of the velocity vector field  around the axis $z$. Soliton
and kink are also the solutions of a boundary problem with free
boundaries.

Let us return to the formulation of the problem. The external helical force $\vec{f}_{0} $ is given in the explicit form in order to make calculations more transparent. Strictly speaking, its explicit form is not very important for the existence itself of $\alpha $-effect. It is necessary that $rot\vec{f}_{0} {\rm \simeq }\vec{f}_{0} $ only. The external force could be chosen statistically by specifying the correlator :
\begin{equation} \label{GEQ225} \overline{f_{i} f_{m} }=A(\tau ,r)\delta _{im} +B(\tau ,r)r_{i} r_{m} +G(\tau ,r)\varepsilon _{imn} r_{n} . \end{equation}
It is fundamental  that the last term $G(\tau ,r)$ (helicity) in this correlator is not equal to zero, otherwise $\alpha $-effect is absent. Nevertheless the statistical method is more bulky since it requires the specification of the functions $A, B, G$ and calculations of rather complicated integrals. If we specify the external force dynamically then the averaging over fast oscillations is performed easily.

The question is of interest about the origin of instability  on qualitative level. For this, we revert to the expression for the Reynolds stresses:
\begin{equation} \label{GEQ226} \nabla _{z} \overline{(v_{0}^{z} v_{2}^{k} +v_{0}^{k} v_{2}^{z} )}. \end{equation}
As far as  the direction $z$ is particular, the averages  $\overline{(v_{0}^{z} v_{2}^{k} +v_{0}^{k} v_{2}^{z} )}$ which are non equal to zero must be proportional to $l_{z} $. From the property of the external force $A_{1} A_{3}^{*} =0$, follows, that with $k=1$, the velocity  $W_{x} $ i fully left of Reynolds stress. Another velocity $W_{y} $ enters into Reynolds stress with the coefficient $iW_{y} l_{z} B_{3} B_{1}^{*} $. Since from the properties of external force it follows that $B_{3} B_{1}^{*} =-\frac{i}{4} $, we obtain the factor $\left(\frac{1}{4} W_{y} l_{z} \right)$. In a similar way, with $k=2$, because of property of the external force $B_{2} B_{3}^{*} =0$, the velocity $W_{y} $ is fully missed out in the second Reynolds equation. The velocity $W_{x} $ enters on the second Reynolds equation with the coefficient: $iW_{x} l_{z} A_{2}^{*} A_{3} $. Due to the property of the external force $A_{2}^{*} A_{3} =\frac{i}{4} $, we obtain the factor $\left(-\frac{1}{4} W_{x} l_{z} \right)$. (We may not write down the common factor). It is clear that as a result we obtain the components of vector product $[\vec{W}\times \vec{l}]=\vec{i}W_{y} l_{z} -\vec{j}W_{x} l_{z} $. Or if we take into account $\nabla _{z} ,$we obtain the components $rot\vec{W}$. These components $rot\vec{W}$ provide a positive feedback loop between  the velocity components like in the  usual $\alpha $-effect which leads to instability.
\begin{figure}
  \centerline{\includegraphics[width=7 cm]{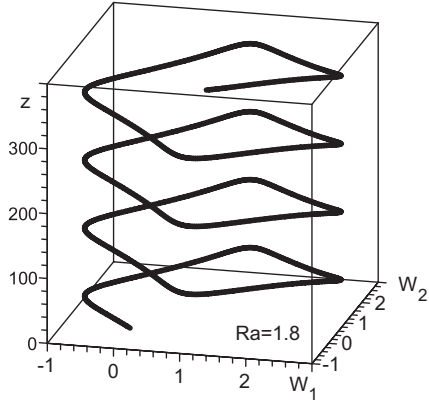}}
  \caption{Helical vortex structure with $Ra=1.8$, $C_1=0$, $C_2=0$.}\label{fig4}
\end{figure}

\section*{Appendix A. Asymptotical development scheme }

In the $R^{-1} $order there is only one term  :

\begin{equation} \label{GEQ227} \partial P_{-1} (X)=0. \end{equation}
With this the equation (\ref{GEQ227}) is satisfied automatically. The equations of the order $R^{0} $ has the form:

\begin{equation} \label{GEQ142} \begin{array}{l} {\partial _{t} v_{0} -\partial ^{2} v_{0}  = -\partial P_{0} +RaT_{0} \vec{l}_{z} +\vec{F}_{0} ,} \\ {\partial _{t} T_{0} -\partial ^{2} T_{0}  = -v_{0}^{z} ,} \\ {\partial v_{0}  = 0.} \end{array} \end{equation}
From the equation  (\ref{GEQ142}) it follows immediately that the functions $v_{0} ,T_{0} ,P_{0} $ are oscillating due to the oscillating character of the external force  $\vec{F}_{0} $. Approximation equations $R^{1} $ have the form:

\begin{equation} \label{GEQ228} \begin{array}{l} {\partial _{t} v_{1} -\partial ^{2} v_{1} +\partial v_{0} v_{0}  = -\partial P_{1} -\nabla P_{-1} (X)+Ra(\Theta _{1} (X)+T_{1} )\vec{l}_{z} ;} \\ {\partial _{t} T_{1} -\partial ^{2} T_{1} = -W_{1}^{z} -v_{1}^{z} -\partial v_{0} T_{0} ;} \\ {\partial v_{1}  = 0} \end{array} \end{equation}
The equations (\ref{GEQ228}) contain already oscillating terms as well as the non-oscillating ones. Oscillating terms after averaging give zero and only  the non-oscillating ones remain. That is why the solvability condition  of this system is the independent vanishing of  the non-oscillating (secular) terms as well as the oscillating ones. For the system (\ref{GEQ228}) the condition of solvability in the approximation $R^{1} $ gives equations:

\begin{equation} \label{GEQ130} \nabla P_{-1} (X)=Ra\Theta _{1} (X)\vec{l}_{z}  \end{equation}

\begin{equation} \label{GEQ129} W_{1}^{z} =0 \end{equation}
The oscillating part  in the approximation $R^{1} $ is described by the equations system:
\begin{equation} \label{GEQ229} \begin{array}{l} {\partial _{t} v_{1} -\partial ^{2} v_{1} +\partial v_{0} v_{0}  = -\partial P_{1} +RaT_{1} \vec{l}_{z} ;} \\ {\partial _{t} T_{1} -\partial ^{2} T_{1}  = -v_{1}^{z} -\partial v_{0} T_{0} ;} \\ {\partial v_{1}  = 0} \end{array} \end{equation}
i.e.$v_{1} =v_{1} (x_{0} ),T_{1} =T_{1} (x_{0} ),P_{1} =P_{1} (x_{0} ).$ In the approximation $R^{2} $ we have the following equations:
\begin{figure}
  \centerline{\includegraphics[width=7 cm]{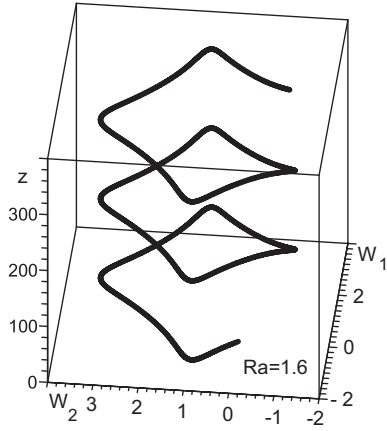}}
  \caption{Example of a vortex structure with $Ra=1.6$, $C_1=0$, $C_2=0$.}\label{fig5}
\end{figure}
\begin{equation} \label{GEQ163} \begin{array}{l} {\partial _{t} v_{2} -\partial ^{2} v_{2} \quad =\quad -\partial P_{2} -\partial (W_{1} v_{0} +v_{0} W_{1} )-\partial (v_{1} v_{0} +v_{0} v_{1} )+RaT_{2} \vec{l}_{z} ;} \\ {\partial _{t} T_{2} -\partial ^{2} T_{2} \quad =\quad -v_{2}^{z} -\partial (W_{1} T_{0} +v_{0} \Theta _{1} )-\partial (v_{1} T_{0} +v_{0} T_{1} );} \\ {\partial v_{2} \quad =\quad 0.} \end{array} \end{equation}
It is easy to evidence that after averaging, all the terms in equations (\ref{GEQ163}) give zero. Thereby secular terms do not appear in the order $R^{2} $ and the fields  $v_{2} ,T_{2} ,P_{2} $ remain oscillating. However, they now depend on large scale variables $X,$ i.e. $v_{2} =v_{2} (x_{0} ,X),T_{2} =T_{2} (x_{0} ,X),$ $P_{2} =P_{2} (x_{0} ,X)$. The approximation $R^{3} $ gives equations:

\begin{equation} \label{GEQ230} \begin{array}{l} {\partial _{t} v_{3} -\partial ^{2} v_{3} \quad =\quad -\partial P_{3} -\partial (W_{1} +v_{1} )(W_{1} +v_{1} )-} \\ {\quad \quad -\partial (v_{2} v_{0} +v_{0} v_{2} )+RaT_{3} \vec{l}_{z} ;} \\ {\partial _{t} T_{3} -\partial ^{2} T_{3} \quad =\quad -v_{3}^{z} -\partial (W_{1} +v_{1} )(\Theta _{1} +T_{1} )-\partial (v_{2} T_{0} +v_{0} T_{2} );} \\ {\partial v_{3} +\nabla W_{1} \quad =\quad 0.} \end{array} \end{equation}
From the equation (\ref{GEQ230}) one can see that an averaging of the first two equations gives only zero terms, i.e. does not give any secular terms. But the third equation averaging gives a new secular term:

\begin{equation} \label{GEQ128} \nabla W_{1} =0 \end{equation}
Thereby the fields $v_{3} ,T_{3} ,P_{3} $ remain oscillating, but depend on $X$, i.e. $v_{3} =v_{3} (x_{0} ,X),T_{3} =T_{3} (x_{0} ,X),$ $P_{3} =P_{3} (x_{0} ,X)$. Now we pass to the equations of the $R^{4} $ approximation. They have the form:
\begin{figure}
  \centerline{\includegraphics[width=7 cm]{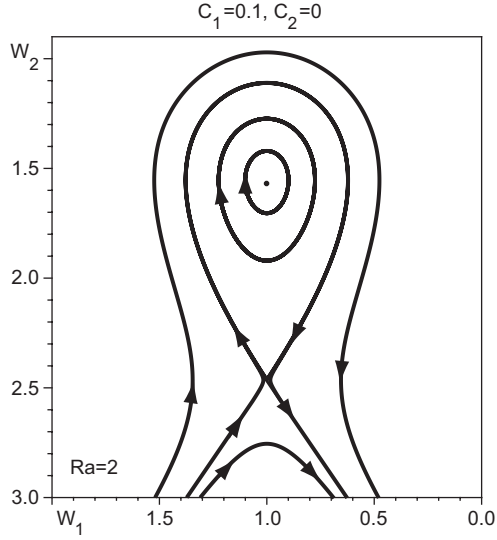}}
  \caption{Phase picture of a dynamical system with $Ra=2$, $C_1=0.1$, $C_2=0$.}\label{fig6}
\end{figure}
\begin{equation} \label{GEQ231} \partial _{t} v_{4} -2\partial \nabla v_{2} -\partial ^{2} v_{4} +\partial [v_{3} v_{0} +v_{0} v_{3} +v_{2} (W_{1} +v_{1} )+(W_{1} +v_{1} )v_{2} ]+ \end{equation}

\[\partial _{t} T_{4} -2\partial \nabla T_{2} -\partial ^{2} T_{4}  = -v_{4}^{z} -\partial \left[v_{3} T_{0} +v_{0} T_{3} +v_{2} (\Theta _{1} +T_{1} )+ \right. \]


\[{\quad \quad +(W_{1} +v_{1} )T_{2} ]-\nabla [(W_{1} +v_{1} )T_{0} +v_{0} (\Theta _{1} +T_{1} )];}\]

\begin{equation}\label{GEQ232}
    {\partial v_{4} +\nabla v_{2}  = 0.}
\end{equation}


It is easy to see that these equations have no secular terms at all.
\begin{figure}
  \centerline{\includegraphics[width=7 cm]{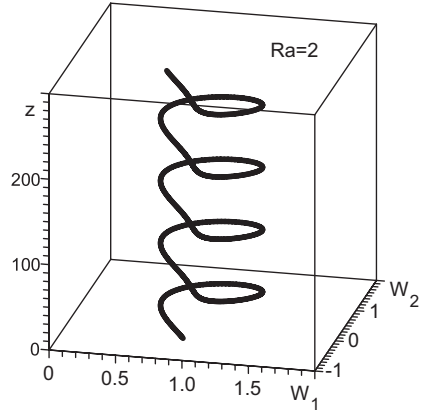}}
  \caption{Helical vortex structure with $Ra=2$, $C_1=0.1$, $C_2=0$. This structure corresponds to the closed trajectory around the elliptical point in fig.\ref{fig6}.}\label{fig7}
\end{figure}

Finally, let us consider equations of the  approximation$R^{5} $. Their form is rather bulky:
\begin{equation} \label{GEQ233} \partial _{t} v_{5} +\partial _{T} W_{1} -\partial ^{2} v_{5} -2\partial \nabla v_{3} -\Delta W_{1} +\partial [v_{4} v_{0} +v_{0} v_{4} +v_{3} (W_{1} +v_{1} )+ \end{equation}

\[+(W_{1} +v_{1} )v_{3} +v_{2} v_{2} ]+ \nabla [ v_{2} v_{0} +v_{0} v_{2} +(W_{1} +v_{1} )(W_{1} +v_{1} )]=-\partial P_{5} -\nabla \overline{P}_{3} (X)-\nabla P_{3} +RaT_{5} \vec{l}_{z} .\]

\begin{equation} \label{GEQ234} \partial _{t} T_{5} +\partial _{T} \Theta _{1} -\partial ^{2} T_{5} -2\partial \nabla T_{3} -\Delta \Theta _{1} = -v_{5}^{z} -\partial [ v_{4} T_{0} +v_{0} T_{4} +v_{3} (\Theta _{1} +T_{1} )+ \end{equation}

\[+(W_{1} +v_{1} )T_{3} +v_{2} T_{2} ]-\nabla (v_{2} T_{0} +v_{0} T_{2} )-\nabla [(W_{1} +v_{1} )(\Theta _{1} +T_{1} )].\]

\[\partial v_{5} +\nabla v_{3} =0\]
The equations of the fifth order (\ref{GEQ233}), (\ref{GEQ234}) give the main system of secular equations  as the solvability conditions of  the fifth approximation:

\begin{equation} \label{GEQ126} \partial _{T} W_{1} -\Delta W_{1} +\nabla \overline{(v_{2} v_{0} +v_{0} v_{2} )}+\nabla (W_{1} W_{1} )=-\nabla \overline{P}_{3} (X); \end{equation}

\begin{equation} \label{GEQ127} \partial _{T} \Theta _{1} -\Delta \Theta _{1} =-\nabla \overline{(v_{2} T_{0} +v_{0} T_{2} )}-\nabla (W_{1} \Theta _{1} ). \end{equation}

\section*{Appendix B. Calculations of the Reynolds stresses.}

Since the external force is composed of four terms (\ref{GEQ161}), then the expression for  $v_{0}^{P} $ is composed of four terms as well: $_{1} v_{0}^{P} ,_{2} v_{0}^{P} ,_{3} v_{0}^{P} ,_{4} v_{0}^{P} $. With this :

\begin{equation} \label{GEQ235} _{2} v_{0}^{P} =(_{1} v_{0}^{P} )^{*} ,_{4} v_{0}^{P} =(_{3} v_{0}^{P} )^{*}  \end{equation}

\begin{equation} \label{GEQ236} _{1} v_{0}^{P} =\exp i\varphi _{1} \left[\delta _{p\mu } -\frac{Ra\widehat{P}_{p\lambda } (k_{1} )l_{\lambda } l_{\mu } }{D_{0}^{2} (-\omega _{0} ,k_{1} )+Ra\widehat{P}_{mq} (k_{1} )l_{m} l_{q} } \right]\frac{A_{\mu } }{D_{0} (-\omega _{0} ,k_{1} )} , \end{equation}

\begin{equation} \label{GEQ237} _{3} v_{0}^{P} =\exp i\varphi _{2} \left[\delta _{p\mu } -\frac{Ra\widehat{P}_{p\lambda } (k_{2} )l_{\lambda } l_{\mu } }{D_{0}^{2} (-\omega _{0} ,k_{2} )+Ra\widehat{P}_{mq} (k_{2} )l_{m} l_{q} } \right]\frac{B_{\mu } }{D_{0} (-\omega _{0} ,k_{2} )} . \end{equation}
\begin{figure}
  \centerline{\includegraphics[width=7 cm]{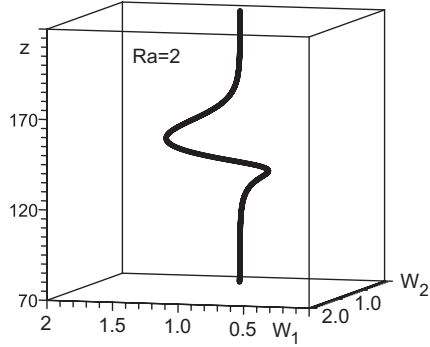}}
  \caption{Helical soliton which corresponds to the separatrix in fig.6 with $Ra=2$, $C_1=0.1$, $C_2=0.1$.}\label{fig8}
\end{figure}
In order to simplify equation writing, we will write down the convolution $\widehat{P}_{mq} l_{m} l_{q} $in the form$\widehat{P}(k)ll$. In the similar way, for the temperature field $T_{0} $ there are four terms: $_{1} T_{0} ,_{2} T_{0} ,_{3} T_{0} ,_{4} T_{0} $. At the same time :

\begin{equation} \label{GEQ238} _{2} T_{0} =(_{1} T_{0} )^{*} ,_{4} T_{0} =(_{3} T_{0} )^{*}  \end{equation}

\begin{equation} \label{GEQ239} _{1} T_{0} =-\exp i\varphi _{1} \left[1-\frac{Ra\widehat{P}(k_{1} )ll}{D_{0}^{2} (-\omega _{0} ,k_{1} )+Ra\widehat{P}(k_{1} )ll} \right]\frac{(l_{\mu } A_{\mu } )}{D_{0}^{2} (-\omega _{0} ,k_{1} )} , \end{equation}

\begin{equation} \label{GEQ240} _{3} T_{0} =-\exp i\varphi _{2} \left[1-\frac{Ra\widehat{P}(k_{2} )ll}{D_{0}^{2} (-\omega _{0} ,k_{2} )+Ra\widehat{P}(k_{2} )ll} \right]\frac{(l_{\mu } B_{\mu } )}{D_{0}^{2} (-\omega _{0} ,k_{2} )}  \end{equation}
Let us consider the tensor $T_{(2)}^{mk} $. It is also composed of four terms: $_{1} T_{(2)}^{mk} ,_{2} T_{(2)}^{mk} ,_{3} T_{(2)}^{mk} ,_{4} T_{(2)}^{mk} $:

\begin{equation} \label{GEQ241} _{2} T_{(2)}^{mk} =(_{1} T_{(2)}^{mk} )^{*} ,_{4} T_{(2)}^{mk} =(_{3} T_{(2)}^{mk} )^{*} . \end{equation}
The equation for the tensor $_{1} T_{(2)}^{mk} $ follows directly from the formula (\ref{GEQ171}):

\begin{equation} \label{GEQ242} \begin{array}{l} {_{1} T_{(2)}^{mk} \quad =\quad \exp i\varphi _{1} \left[\delta _{kj} -\frac{Ra\widehat{P}_{kn} (k_{1} )l_{n} l_{j} }{D_{0}^{2} (-\omega _{0} ,k_{1} )+Ra\widehat{P}(k_{1} )ll} \right]\frac{\widehat{P}_{jp} (k_{1} )ik^{m} }{D_{0}^{2} (-\omega _{0} ,k_{1} )} \times } \\ {\quad \quad \times \left[\delta _{p\mu } -\frac{Ra\widehat{P}_{p\lambda } (k_{1} )l_{\lambda } l_{\mu } }{D_{0}^{2} (-\omega _{0} ,k_{1} )+Ra\widehat{P}(k_{1} )ll} \right]A_{\mu } .} \end{array} \end{equation}
First of all we transform the expression, which contains
projection operators in the formula (\ref{GEQ242}).

\[\left[\delta _{kj} -\frac{Ra\widehat{P}_{kn} (k_{1} )l_{n} l_{j} }{D_{0}^{2} (-\omega _{0} ,k_{1} )+Ra\widehat{P}(k_{1} )ll} \right]\widehat{P}_{jp} \left[\delta _{p\mu } -\frac{Ra\widehat{P}_{p\lambda } (k_{1} )l_{\lambda } l_{\mu } }{D_{0}^{2} (-\omega _{0} ,k_{1} )+Ra\widehat{P}(k_{1} )ll} \right]=\]

\begin{equation} \label{GEQ243} \left[\delta _{kj} -\frac{Ra\widehat{P}_{kn} (k)l_{n} l_{j} }{D_{0}^{2} (\omega _{0} ,k)-Ra\widehat{P}(k)ll} \right]\widehat{P}_{j\lambda } \left[\delta _{\lambda \mu } -\frac{Ral_{\lambda } l_{\mu } }{D_{0}^{2} (\omega _{0} ,k)-Ra\widehat{P}(k)ll} \right]= \end{equation}

\[=\widehat{P}_{kn} \left[\delta _{n\lambda } -\frac{Ra\widehat{P}_{\lambda j} (k)l_{n} l_{j} }{D_{0}^{2} (\omega _{0} ,k)+Ra\widehat{P}(k)ll} \right]\left[\delta _{\lambda \mu } -\frac{Ral_{\lambda } l_{\mu } }{D_{0}^{2} (\omega _{0} ,k)+Ra\widehat{P}(k)ll} \right].\]
Here we use the property of the projection operator.
\begin{figure}
  \centerline{\includegraphics[width=7 cm]{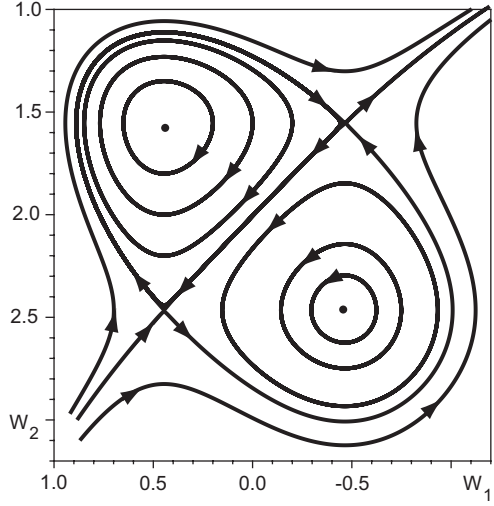}}
  \caption{Phase picture of the dynamical system with $Ra=2$, $C_1=0.1$, $C_2=0.1$. One can see the appearance of two hyperbolical and two elliptical points.}\label{fig9}
\end{figure}
\[\widehat{P}_{jp} \widehat{P}_{p\lambda } =\widehat{P}_{j\lambda } \]
With help of the formulae (\ref{GEQ243}) it is possible to write down:

\begin{equation} \label{GEQ244} _{1} T_{(2)}^{mk} =\exp i\varphi _{1} \frac{ik_{1}^{m} \widehat{P}_{kn} (k_{1} )}{D_{0}^{2} (-\omega _{0} ,k_{1} )} \left[\delta _{n\lambda } -\frac{Ra\widehat{P}_{\lambda j} (k_{1} )l_{n} l_{j} }{D_{0}^{2} (-\omega _{0} ,k_{1} )+Ra\widehat{P}(k_{1} )ll} \right]\times  \end{equation}

\[\times \left[\delta _{\lambda \mu } -\frac{Ral_{\lambda } l_{\mu } }{D_{0}^{2} (-\omega _{0} ,k_{1} )+Ra\widehat{P}(k_{1} )ll} \right]A_{\mu } ,\]

\begin{equation} \label{GEQ245} _{3} T_{(2)}^{mk} =\exp i\varphi _{2} \frac{ik_{2}^{m} \widehat{P}_{kn} (k_{2} )}{D_{0}^{2} (-\omega _{0} ,k_{2} )} \left[\delta _{n\lambda } -\frac{Ra\widehat{P}_{\lambda j} (k_{2} )l_{n} l_{j} }{D_{0}^{2} (-\omega _{0} ,k_{2} )+Ra\widehat{P}(k_{2} )ll} \right]\times  \end{equation}

\[\times \left[\delta _{\lambda \mu } -\frac{Ral_{\lambda } l_{\mu } }{D_{0}^{2} (-\omega _{0} ,k_{2} )+Ra\widehat{P}(k_{2} )ll} \right]B_{\mu } .\]
From the definition (\ref{GEQ170}) follows expressions for four terms of the tensor:
$T_{(1)}^{mk} $:$_{1} T_{(1)}^{mk} ,_{2} T_{(1)}^{mk} ,_{3} T_{(1)}^{mk} ,_{4} T_{(1)}^{mk} :$

\begin{equation} \label{GEQ246} _{2} T_{(1)}^{mk} =(_{1} T_{(1)}^{mk} )^{*} ,_{4} T_{(1)}^{mk} =(_{3} T_{(1)}^{mk} )^{*} . \end{equation}

\begin{equation} \label{GEQ247} _{1} T_{(1)}^{mk} =-\exp i\varphi _{1} \frac{ik_{1}^{m} Ral^{n} l^{\mu } }{D_{0}^{4} (-\omega _{0} ,k_{1} )} \widehat{P}_{kn} (k_{1} )\times  \end{equation}

\[\times \left[1-\frac{Ra\widehat{P}(k_{1} )ll}{D_{0}^{2} (-\omega _{0} ,k_{1} )+Ra\widehat{P}(k_{1} )ll} \right]^{2} A_{\mu } ,\]

\begin{equation} \label{GEQ248} _{3} T_{(1)}^{mk} =-\exp i\varphi _{2} \frac{ik_{2}^{m} Ral^{n} l^{\mu } }{D_{0}^{4} (-\omega _{0} ,k_{2} )} \widehat{P}_{kn} (k_{1} )\times  \end{equation}

\[\times \left[1-\frac{Ra\widehat{P}(k_{2} )ll}{D_{0}^{2} (-\omega _{0} ,k_{2} )+Ra\widehat{P}(k_{2} )ll} \right]^{2} B_{\mu } .\]
As a matter of fact, all these expressions are considerably simplified. Actually, the operator $D_{0} =\partial _{t} -\partial ^{2} $ acts on its own function $\exp (i\omega _{0} t+i\vec{k}\vec{x})$ and gives $D_{0} \exp (i\omega _{0} t+i\vec{k}\vec{x})=D_{0} (\omega _{0} ,k_{0} )\exp (i\omega _{0} t+i\vec{k}\vec{x})$, where $D_{0} (\omega _{0} ,k_{0} )=i\omega _{0} +k_{0}^{2} $. In dimensionless variables this means that  :

\begin{equation} \label{GEQ249} D_{0} (\omega _{0} ,k_{0} )=1+i;D_{0} (-\omega _{0} ,k_{0} )=1-i=D_{0}^{*} (\omega _{0} ,k_{0} ). \end{equation}
The expression $\widehat{P}_{kn} l_{k} l_{n} \equiv \widehat{P}ll=l^{2} -\frac{(kl)^{2} }{k^{2} } =1,$ since $\vec{k}_{1} ,\vec{k}_{2} \bot \vec{l}$. As a result:
\begin{figure}
  \centerline{\includegraphics[width=7 cm]{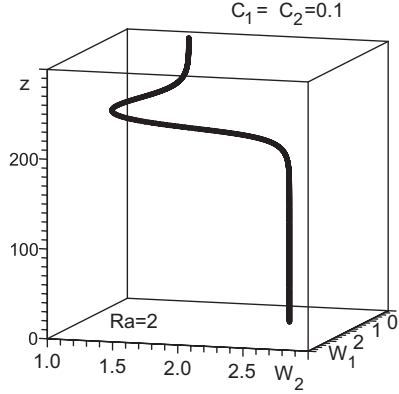}}
  \caption{Helical kink which corresponds to the separatrix in fig.\ref{fig9}.}\label{fig10}
\end{figure}
\begin{equation} \label{GEQ250} D_{0}^{2} (\omega _{0} ,k_{0} )+Ra\widehat{P}ll=2i+Ra;D_{0}^{2} (-\omega _{0} ,k_{0} )+Ra\widehat{P}ll=(2i+Ra)^{*} . \end{equation}
And all tensors are simplified considerably:
\begin{equation} \label{GEQ251} _{1} v_{0}^{P} =\exp i\varphi _{1} \left[\delta _{pm} -\frac{Ra\widehat{P}_{p\nu } (k_{1} )l_{\nu } l_{m} }{(2i+Ra)^{*} } \right]\frac{A_{m} }{(1-i)} , \end{equation}

\begin{equation} \label{GEQ252} _{3} v_{0}^{P} =\exp i\varphi _{2} \left[\delta _{pm} -\frac{Ra\widehat{P}_{p\nu } (k_{2} )l_{\nu } l_{m} }{(2i+Ra)^{*} } \right]\frac{B_{m} }{(1-i)} . \end{equation}

\begin{equation} \label{GEQ253} _{1} T_{(2)}^{mk} =\exp i\varphi _{1} \frac{ik_{1}^{m} \widehat{P}_{kn} (k_{1} )}{(1-i)^{2} } \left[\delta _{n\lambda } -\frac{Ra\widehat{P}_{\lambda j} (k_{1} )l_{n} l_{j} }{(2i+Ra)^{*} } \right]\times  \end{equation}

\[\times \left[\delta _{\lambda \mu } -\frac{Ral_{\lambda } l_{\mu } }{(2i+Ra)^{*} } \right]A_{\mu } ,\]

\begin{equation} \label{GEQ254} _{3} T_{(2)}^{mk} =\exp i\varphi _{2} \frac{ik_{2}^{m} \widehat{P}_{kn} (k_{2} )}{(1-i)^{2} } \left[\delta _{n\lambda } -\frac{Ra\widehat{P}_{\lambda j} (k_{2} )l_{n} l_{j} }{(2i+Ra)^{*} } \right]\times  \end{equation}

\[\times \left[\delta _{\lambda \mu } -\frac{Ral_{\lambda } l_{\mu } }{(2i+Ra)^{*} } \right]B_{\mu } .\]

\begin{equation} \label{GEQ255} _{1} T_{(1)}^{mk} =-\exp i\varphi _{1} \frac{ik_{1}^{m} Ral^{n} l^{\mu } }{(1-i)^{4} } \widehat{P}_{kn} (k_{1} )\times  \end{equation}

\[\times \left[1-\frac{Ra}{(2i+Ra)^{*} } \right]^{2} A_{\mu } ,\]

\begin{equation} \label{GEQ256} _{3} T_{(1)}^{mk} =-\exp i\varphi _{2} \frac{ik_{2}^{m} Ral^{n} l^{\mu } }{(1-i)^{4} } \widehat{P}_{kn} (k_{1} )\times  \end{equation}

\[\times \left[1-\frac{Ra}{(2i+Ra)^{*} } \right]^{2} B_{\mu } .\]
The reason of the further simplification of these expressions is :

\begin{equation} \label{GEQ257} \widehat{P}_{p\nu } (k)l_{\nu } =l_{p} . \end{equation}
Taking into account (\ref{GEQ257}), we obtain:
\begin{equation} \label{GEQ258} _{1} v_{0}^{P} =\exp i\varphi _{1} \left[\delta _{pm} -\frac{Ral_{p} l_{m} }{(2i+Ra)^{*} } \right]\frac{A_{m} }{(1-i)} , \end{equation}

\begin{equation} \label{GEQ259} _{3} v_{0}^{P} =\exp i\varphi _{2} \left[\delta _{pm} -\frac{Ral_{p} l_{m} }{(2i+Ra)^{*} } \right]\frac{B_{m} }{(1-i)} . \end{equation}

\begin{equation} \label{GEQ260} _{1} T_{(2)}^{mk} =\exp i\varphi _{1} \frac{ik_{1}^{m} \widehat{P}_{kn} (k_{1} )}{(1-i)^{2} } \left[\delta _{n\lambda } -\frac{Ral_{n} l_{\lambda } }{(2i+Ra)^{*} } \right]\times  \end{equation}

\[\times \left[\delta _{\lambda \mu } -\frac{Ral_{\lambda } l_{\mu } }{(2i+Ra)^{*} } \right]A_{\mu } ,\]

\begin{equation} \label{GEQ261} _{3} T_{(2)}^{mk} =\exp i\varphi _{2} \frac{ik_{2}^{m} \widehat{P}_{kn} (k_{2} )}{(1-i)^{2} } \left[\delta _{n\lambda } -\frac{Ral_{n} l_{\lambda } }{(2i+Ra)^{*} } \right]\times . \end{equation}

\[\times \left[\delta _{\lambda \mu } -\frac{Ral_{\lambda } l_{\mu } }{(2i+Ra)^{*} } \right]B_{\mu } .\]

\begin{equation} \label{GEQ262} _{1} T_{(1)}^{mk} =-\exp i\varphi _{1} \frac{ik_{1}^{m} Ral^{k} l^{\mu } }{(1-i)^{4} } \left[1-\frac{Ra}{(2i+Ra)^{*} } \right]^{2} A_{\mu } , \end{equation}

\begin{equation} \label{GEQ263} _{3} T_{(1)}^{mk} =-\exp i\varphi _{2} \frac{ik_{2}^{m} Ral^{k} l^{\mu } }{(1-i)^{4} } \left[1-\frac{Ra}{(2i+Ra)^{*} } \right]^{2} B_{\mu } . \end{equation}
We will do calculations of  Reynolds stresses in several stages. To begin with, we consider the term $\overline{v_{0}^{p} T_{(2)}^{mk} }$. This average value is composed of four terms in which the oscillation phase is cancelled:

\begin{equation} \label{GEQ264} \overline{v_{0}^{p} T_{(2)}^{mk} }=\overline{[_{1} v_{0}^{P} ][_{2} T_{(2)}^{mk} ]}+\overline{[_{2} v_{0}^{P} ][_{1} T_{(2)}^{mk} ]}+\overline{[_{3} v_{0}^{P} ][_{4} T_{(2)}^{mk} ]}+\overline{[_{4} v_{0}^{P} ][_{3} T_{(2)}^{mk} ]} \end{equation}
The second term in the  (\ref{GEQ264}) is conjugated with the first one, and the fourth with the third one. Now we substitute in the (\ref{GEQ264}) correspondent expressions for tensors and we obtain:

\begin{equation} \label{GEQ265} \overline{v_{0}^{p} T_{(2)}^{mk} }=Re\frac{(-ik_{1}^{m} )}{(1+i)} \left[A^{p} -\frac{Ral^{p} A^{z} }{(2i+Ra)^{*} } \right]\times  \end{equation}

\[\times \left[ A_{k}^{*} -k_{1}^{k} A_{x}^{*} -\frac{2Ral_{k} A_{z}^{*} }{(2i+Ra)} +\frac{Ra^{2} l_{k} A_{z}^{*} }{(2i+Ra)^{2} } \right]+\]

\[+Re\left[\frac{(-ik_{2}^{m} )}{(1+i)} [B^{p} -\frac{Ral^{p} B^{z} }{(2i+Ra)^{*} } ][B_{k}^{*} -k_{2}^{k} B_{y}^{*} -\frac{2Ral_{k} B_{z}^{*} }{(2i+Ra)} +\frac{Ra^{2} l_{k} B_{z}^{*} }{(2i+Ra)^{2} } ]\right].\]
Let us find now the components of Reynolds stresses  $\overline{v_{0}^{p} v_{2}^{k} }=$ $-\overline{v_{0}^{p} W_{1}^{m} T_{(2)}^{mk} }$. Taking in consideration the (\ref{GEQ265}), we obtain:

\begin{equation} \label{GEQ266} \; \overline{v_{0}^{p} v_{2}^{k} }=Re\frac{iW_{x} }{(1+i)} \left[A^{p} -\frac{Ral^{p} A^{z} }{(2i+Ra)^{*} } \right]\times  \end{equation}

\[\times \left[ A_{k}^{*} -k_{1}^{k} A_{x}^{*} -\frac{2Ral_{k} A_{z}^{*} }{(2i+Ra)} +\frac{Ra^{2} l_{k} A_{z}^{*} }{(2i+Ra)^{2} } \right]+\]

\[+Re\left[\frac{iW_{y} }{(1+i)} [B^{p} -\frac{Ral^{p} B^{z} }{(2i+Ra)^{*} } ][B_{k}^{*} -k_{2}^{k} B_{y}^{*} -\frac{2Ral_{k} B_{z}^{*} }{(2i+Ra)} +\frac{Ra^{2} l_{k} B_{z}^{*} }{(2i+Ra)^{2} } ]\right]\]
The full contribution in the tensor of the Reynolds stresses  $\overline{v_{0}^{p} v_{2}^{k} }+\overline{v_{0}^{k} v_{2}^{p} }$ from the tensor $T_{(2)}^{mk} $ is obtained using the symmetrization of this equation over the indices  $p,k$. As a result we obtain:
\begin{equation} \label{GEQ267} \overline{v_{0}^{p} v_{2}^{k} }+\overline{v_{0}^{k} v_{2}^{p} }=Re\frac{iW_{x} }{(1+i)} \left[A^{p} -\frac{Ral^{p} A^{z} }{(2i+Ra)^{*} } \right]\times  \end{equation}

\[\times \left[ A_{k}^{*} -k_{1}^{k} A_{x}^{*} -\frac{2Ral_{k} A_{z}^{*} }{(2i+Ra)} +\frac{Ra^{2} l_{k} A_{z}^{*} }{(2i+Ra)^{2} } \right]+\]

\[+Re\left[\frac{iW_{y} }{(1+i)} [B^{p} -\frac{Ral^{p} B^{z} }{(2i+Ra)^{*} } ][B_{k}^{*} -k_{2}^{k} B_{y}^{*} -\frac{2Ral_{k} B_{z}^{*} }{(2i+Ra)} +\frac{Ra^{2} l_{k} B_{z}^{*} }{(2i+Ra)^{2} } ]\right]+\]

\[Re\left[\frac{iW_{x} }{(1+i)} [A^{k} -\frac{Ral^{k} A^{z} }{(2i+Ra)^{*} } ][A_{p}^{*} -k_{1}^{p} A_{x}^{*} -\frac{2Ral_{p} A_{z}^{*} }{(2i+Ra)} +\frac{Ra^{2} l_{p} A_{z}^{*} }{(2i+Ra)^{2} } ]\right]+\]

\[+Re\left[\frac{iW_{y} }{(1+i)} [B^{k} -\frac{Ral^{k} B^{z} }{(2i+Ra)^{*} } ][B_{p}^{*} -k_{2}^{p} B_{y}^{*} -\frac{2Ral_{p} B_{z}^{*} }{(2i+Ra)} +\frac{Ra^{2} l_{p} B_{z}^{*} }{(2i+Ra)^{2} } ]\right].\]
Let us put $p=z,k=1,$ in the equation (\ref{GEQ267}),  i.e. we find the component $x$ in the Reynolds stress. It is easy to see that:

\begin{equation} \label{GEQ268} \overline{v_{0}^{z} v_{2}^{1} }+\overline{v_{0}^{1} v_{2}^{z} }=Re\nabla _{z} \left\{\frac{iW_{y} }{(1+i)} [1-\frac{Ra}{(2i+Ra)^{*} } ]B^{z} B_{1}^{*} \right\}+ \end{equation}

\[+Re\nabla _{z} \left\{\frac{iW_{y} }{(1+i)} [1-\frac{2Ra}{(2i+Ra)} +\frac{Ra^{2} }{(2i+Ra)^{2} } ]B_{z}^{*} B_{1} \right\}.\]
Taking into account the $B_{z}^{*} B_{1} =\frac{i}{4} $, we obtain:

\begin{equation} \label{GEQ269} \overline{v_{0}^{z} v_{2}^{1} }+\overline{v_{0}^{1} v_{2}^{z} }= \end{equation}

\[=Re\nabla _{z} W_{y} \left\{\frac{1}{4(1+i)} \left[-\frac{Ra}{(2i+Ra)^{*} } +\frac{2Ra}{(2i+Ra)} -\frac{Ra^{2} }{(2i+Ra)^{2} } \right]\right\}.\]
After calculating the real part in the  (\ref{GEQ269}) we obtain:
\begin{equation} \label{GEQ270} \overline{v_{0}^{z} v_{2}^{1} }+\overline{v_{0}^{1} v_{2}^{z} }=\alpha _{1} \nabla _{z} W_{y} , \end{equation}
Where :
\begin{equation} \label{GEQ271} \alpha _{1} =-Ra\frac{(-4Ra+Ra^{2} +12)}{4(Ra^{2} +4)^{2} } . \end{equation}
Putting $p=z,k=2,$ in the (\ref{GEQ267}), we find the corresponding component in the Reynolds stress. It is easy to see that :
\begin{equation} \label{GEQ272} \overline{v_{0}^{z} v_{2}^{2} }+\overline{v_{0}^{2} v_{2}^{z} }=Re\nabla _{z} \frac{iW_{x} }{(1+i)} \times  \end{equation}

\[\times \left[-\frac{Ra}{(2i+Ra)^{*} } A_{3} A_{2}^{*} -\frac{2Ra}{(2i+Ra)} A_{2} A_{3}^{*} +\frac{Ra^{2} }{(2i+Ra)^{2} } A_{2} A_{3}^{*} \right]+\]

\[+Re\left\{\nabla _{z} \frac{iW_{y} }{(1+i)} \left[1-\frac{2Ra}{(2i+Ra)} +\frac{Ra^{2} }{(2i+Ra)^{2} } \right]B_{2} B_{3}^{*} \right\}.\]

As far as  $B_{2} B_{3}^{*} =0,A_{3} A_{2}^{*} =\frac{i}{4} $, we obtain from the (\ref{GEQ272}):

\begin{equation} \label{GEQ273} \overline{v_{0}^{z} v_{2}^{2} }+\overline{v_{0}^{2} v_{2}^{z} }=-\alpha _{1} \nabla _{z} W_{x} . \end{equation}

Now we need to calculate the contribution in the Reynolds stresses from the tensor $T_{(1)}^{mk} $. As it was done previously

\begin{equation} \label{GEQ274} \overline{v_{0}^{p} T_{(1)}^{mk} }=\overline{[_{1} v_{0}^{P} ][_{2} T_{(1)}^{mk} ]}+\overline{[_{2} v_{0}^{P} ][_{1} T_{(1)}^{mk} ]}+\overline{[_{3} v_{0}^{P} ][_{4} T_{(1)}^{mk} ]}+\overline{[_{4} v_{0}^{P} ][_{3} T_{(1)}^{mk} ]} \end{equation}
With this the second term is conjugated with the first one and the fourth is conjugated with the third one. The simple calculation of the first term gives:
\begin{equation} \label{GEQ275} \overline{[_{1} v_{0}^{P} ][_{2} T_{(1)}^{mk} ]}=-\frac{Ra}{2(1+i)^{3} } \left(1-\frac{Ra}{(2i+Ra)} \right)^{2} (-ik_{1}^{m} )l^{k} \times  \end{equation}

\[\times \left[A_{p} -\frac{Ral_{p} A_{3} }{(2i+Ra)^{*} } \right]A_{3}^{*} .\]
We calculate similarly the third term:
\begin{equation} \label{GEQ276} \overline{[_{3} v_{0}^{P} ][_{4} T_{(1)}^{mk} ]}=-\frac{Ra}{2(1+i)^{3} } \left(1-\frac{Ra}{(2i+Ra)} \right)^{2} (-ik_{2}^{m} )l^{k} \times  \end{equation}

\[\times \left[B_{p} -\frac{Ral_{p} B_{3} }{(2i+Ra)^{*} } \right]B_{3}^{*} \]
Now it is easy to find the contribution in the $\overline{v_{0}^{p} v_{2}^{k} }$:

\begin{equation} \label{GEQ277} \overline{v_{0}^{p} v_{2}^{k} }= \end{equation}
\[=Re\left\{-\frac{Ra}{(1+i)^{3} } \left(1-\frac{Ra}{(2i+Ra)} \right)^{2} (iW_{x} )l^{k} \left[A_{p} -\frac{Ral_{p} A_{3} }{(2i+Ra)^{*} } \right]A_{3}^{*} \right\}+\]

\[+Re\left\{-\frac{Ra}{(1+i)^{3} } \left(1-\frac{Ra}{(2i+Ra)} \right)^{2} (iW_{y} )l^{k} \left[B_{p} -\frac{Ral_{p} B_{3} }{(2i+Ra)^{*} } \right]B_{3}^{*} \right\}.\]
After symmetrizing this tensor over the indices  $p,k$, we obtain:
\begin{equation} \label{GEQ278} \overline{v_{0}^{p} v_{2}^{k} }+\overline{v_{0}^{k} v_{2}^{p} }= \end{equation}

\[=Re\left\{-\frac{Ra}{(1+i)^{3} } \left(1-\frac{Ra}{(2i+Ra)} \right)^{2} (iW_{x} )l^{k} \left[A_{p} -\frac{Ral_{p} A_{3} }{(2i+Ra)^{*} } \right]A_{3}^{*} \right\}+\]

\[+Re\left\{-\frac{Ra}{(1+i)^{3} } \left(1-\frac{Ra}{(2i+Ra)} \right)^{2} (iW_{y} )l^{k} \left[B_{p} -\frac{Ral_{p} B_{3} }{(2i+Ra)^{*} } \right]B_{3}^{*} \right\}+\]

\[+ Re\left\{-\frac{Ra}{(1+i)^{3} } \left(1-\frac{Ra}{(2i+Ra)} \right)^{2} (iW_{x} )l^{p} \left[A_{k} -\frac{Ral_{k} A_{3} }{(2i+Ra)^{*} } \right]A_{3}^{*} \right\}+\]

\[+Re\left\{-\frac{Ra}{(1+i)^{3} } \left(1-\frac{Ra}{(2i+Ra)} \right)^{2} (iW_{y} )l^{p} \left[B_{k} -\frac{Ral_{k} B_{3} }{(2i+Ra)^{*} } \right]B_{3}^{*} \right\}.\]
Putting $p=z,k=1$, in the formula(\ref{GEQ278}), we obtain the tensor $x$-component of the Reynolds stress:

\begin{equation} \label{GEQ279} \overline{v_{0}^{z} v_{2}^{1} }+\overline{v_{0}^{1} v_{2}^{z} }= \end{equation}

\[=Re\left\{-\frac{iRa}{(1+i)^{3} } \left(1-\frac{Ra}{(2i+Ra)} \right)^{2} \nabla _{z} \left[W_{x} A_{1} A_{3}^{*} +W_{y} B_{1} B_{3}^{*} \right]\right\}.\]
Since $A_{1} A_{3}^{*} =0,B_{1} B_{3}^{*} =\frac{i}{4} $, we get:

\begin{equation} \label{GEQ280} \overline{v_{0}^{z} v_{2}^{1} }+\overline{v_{0}^{1} v_{2}^{z} }=Re\left\{\frac{Ra}{4(1+i)^{3} } \left(1-\frac{Ra}{(2i+Ra)} \right)^{2} \nabla _{z} W_{y} \right\} \end{equation}
After calculating the real part, we obtain:
\begin{equation} \label{GEQ281} \overline{v_{0}^{z} v_{2}^{1} }+\overline{v_{0}^{1} v_{2}^{z} }=\alpha _{2} \nabla _{z} W_{y}  \end{equation}
Where

\begin{equation} \label{GEQ282} \alpha _{2} =-Ra\frac{\left(4-Ra^{2} -4Ra\right)}{4(4+Ra^{2} )^{2} } . \end{equation}
Putting in the formula  (\ref{GEQ278}) $p=z,k=2$ we easily get:

\begin{equation} \label{GEQ283} \overline{v_{0}^{z} v_{2}^{2} }+\overline{v_{0}^{2} v_{2}^{z} }= \end{equation}

\[=Re\left\{-\frac{iRa}{(1+i)^{3} } \left(1-\frac{Ra}{(2i+Ra)} \right)^{2} \nabla _{z} \left[W_{x} A_{2} A_{3}^{*} +W_{y} B_{2} B_{3}^{*} \right]\right\}.\]
Since $A_{2} A_{3}^{*} =-\frac{i}{4} ,B_{2} B_{3}^{*} =0,$then

\begin{equation} \label{GEQ284} \overline{v_{0}^{z} v_{2}^{2} }+\overline{v_{0}^{2} v_{2}^{z} }=-\alpha _{2} \nabla _{z} W_{x}  \end{equation}
and designating $:$

\begin{equation} \label{GEQ285} \alpha =\alpha _{1} +\alpha _{2} =-Ra\frac{4-2Ra}{(4+Ra^{2} )^{2} } , \end{equation}
we obtain the final expressions for the tensor components of the Reynolds stresses:
\begin{equation} \label{GEQ286} \overline{v_{0}^{z} v_{2}^{1} }+\overline{v_{0}^{1} v_{2}^{z} }=\alpha \nabla _{z} W_{y} , \end{equation}

\begin{equation} \label{GEQ287} \overline{v_{0}^{z} v_{2}^{2} }+\overline{v_{0}^{2} v_{2}^{z} }=-\alpha \nabla _{z} W_{x} . \end{equation}
The terms (\ref{GEQ286}) and (\ref{GEQ287}) are fundamental. Nevertheless, strictly speaking one must calculate other, less important, terms. Let us consider the component $p=z,k=3$ of the Reynolds stress. Putting in (\ref{GEQ267}) $p=z,k=3$, we get the contribution of the tensor $T_{(2)}^{mk} $:

\begin{equation} \label{GEQ288} \overline{v_{0}^{z} v_{2}^{3} }+\overline{v_{0}^{3} v_{2}^{z} }= \end{equation}

\[=Re\nabla _{z} \left\{\frac{2iW_{x} }{1+i} \left[1-\frac{Ra}{(2i+Ra)^{*} } \right]\left[1-\frac{2Ra}{(2i+Ra)} +\frac{Ra^{2} }{(2i+Ra)^{2} } \right]A_{3} A_{3}^{*} \right\}+\]

\[+Re\nabla _{z} \left\{\frac{2iW_{y} }{1+i} \left[1-\frac{Ra}{(2i+Ra)^{*} } \right]\left[1-\frac{2Ra}{(2i+Ra)} +\frac{Ra^{2} }{(2i+Ra)^{2} } \right]B_{3} B_{3}^{*} \right\}.\]
As far as  $A_{3} A_{3}^{*} =B_{3} B_{3}^{*} =\frac{1}{4} $, we obtain:
\begin{equation} \label{GEQ289} \overline{v_{0}^{z} v_{2}^{3} }+\overline{v_{0}^{3} v_{2}^{z} }=C_{2} \nabla _{z} \left(W_{x} +W_{y} \right), \end{equation}
where
\begin{equation} \label{GEQ290} C_{2} =\frac{-Ra^{3} +2Ra^{2} -4Ra+8}{(4+Ra^{2} )^{3} } . \end{equation}
Now we can find the contribution in $\overline{v_{0}^{z} v_{2}^{3}
}+\overline{v_{0}^{3} v_{2}^{z} }$ of the tensor $T_{(1)}^{mk} $.
Putting in the formula (\ref{GEQ278}) $p=z,k=3$, we obtain:
\begin{equation} \label{GEQ291} \overline{v_{0}^{z} v_{2}^{3} }+\overline{v_{0}^{3} v_{2}^{z} }=Re\nabla _{z} \left\{\frac{1}{2} \left[1-\frac{Ra}{(2i+Ra)^{*} } \right]\left(W_{x} +W_{y} \right)\right\}, \end{equation}
\begin{equation} \label{GEQ292} \overline{v_{0}^{z} v_{2}^{3} }+\overline{v_{0}^{3} v_{2}^{z} }=C_{1} \nabla _{z} \left(W_{x} +W_{y} \right), \end{equation}
where

\begin{equation} \label{GEQ293} C_{1} =-Ra\frac{2+Ra}{(4+Ra^{2} )^{2} }  \end{equation}
Designating the common coefficient $C=C_{1} +C_{2} ,$

\begin{equation} \label{GEQ294} C=\frac{8-12Ra-2Ra^{2} -3Ra^{3} -Ra^{4} }{(4+Ra^{2} )^{3} } , \end{equation}
we obtain:

\begin{equation} \label{GEQ295} \overline{v_{0}^{z} v_{2}^{3} }+\overline{v_{0}^{3} v_{2}^{z} }=C\nabla _{z} \left(W_{x} +W_{y} \right), \end{equation}

\section{Appendix C. Closure of the temperature equation.}

In order to close the temperature equation we have to calculate the term:
\begin{equation} \label{GEQ296} \nabla _{z} \left(\overline{v_{(2)}^{z} T_{(0)} }+\overline{v_{(0)}^{z} T_{(2)} }\right) \end{equation}
It follows from the formula (\ref{GEQ165}), that indeed there is the contribution only of the following terms in $T_{(2)} $:

\begin{equation} \label{GEQ297} T_{(2)} =-\frac{1}{D_{0} } v_{(2)}^{z} -\frac{1}{D_{0} } W_{p} \partial _{p} T_{(0)} . \end{equation}
From the (\ref{GEQ156}) it is easy to find the component $_{1} T_{(0)} ,_{2} T_{(0)} ,_{3} T_{(0)} ,_{4} T_{(0)} ,$
$_{1} T_{(0)}^{*} ={_{2} T_{(0)} }$, $_{3} T_{(0)}^{*} =_{4} T_{(0)}$ .

\begin{equation} \label{GEQ298} _{2} T_{(0)} =-\exp (-i\varphi _{1} )\frac{1}{(1+i)^{2} } \left[1-\frac{Ra}{(2i+Ra)} \right]A_{3}^{*} , \end{equation}

\begin{equation} \label{GEQ299} _{4} T_{(0)} =-\exp (-i\varphi _{2} )\frac{1}{(1+i)^{2} } \left[1-\frac{Ra}{(2i+Ra)} \right]B_{3}^{*} . \end{equation}

\begin{equation} \label{GEQ300} T_{(2)} =W_{1}^{p} T_{(1)}^{pz} +W_{1}^{p} T_{(2)}^{pz} T-\frac{1}{D_{0} } W_{p} \partial _{p} T_{(0)}  \end{equation}
First of all we find the component $\nabla _{z} \overline{v_{(2)}^{z} T_{(0)} }:$

\begin{equation} \label{GEQ301} \nabla _{z} \overline{v_{(2)}^{z} T_{(0)} }=2Re\nabla _{z} \left\{\left[_{1} v_{(2)}^{z} \right]\left[_{2} T_{(0)} \right]+\left[_{3} v_{(2)}^{z} \right]\left[_{4} T_{(0)} \right]\right\}. \end{equation}
Since the $v_{(2)}^{z} $ is composed of two terms (\ref{GEQ169}), then at the beginning we find the contribution -$W^{q} [_{1} T_{(1)}^{qz} ][_{2} T_{(0)} ]$. As it was done in previous calculations we obtain:

\begin{equation} \label{GEQ302} -W^{q} [_{1} T_{(1)}^{qz} ][_{2} T_{(0)} ]= \end{equation}

\[=\frac{(-i)RaW_{x} }{(1-i)^{4} (1+i)^{2} } \left[1-\frac{Ra}{(2i+Ra)^{*} } \right]\left|1-\frac{Ra}{(2i+Ra)^{*} } \right|^{2} A_{z} A_{z}^{*} .\]
Further we  find the contribution:

\begin{equation} \label{GEQ303} -W^{q} [_{1} T_{(2)}^{qz} ][_{2} T_{(0)} ]=\frac{iW_{x} }{4} \left[1-\frac{Ra}{(2i+Ra)^{*} } \right]\left|1-\frac{Ra}{(2i+Ra)} \right|^{2} A_{z} A_{z}^{*} . \end{equation}
As a result we get the term:

\begin{equation} \label{GEQ304} 2Re\nabla _{z} \left[_{1} v_{(2)}^{z} \right]\left[_{2} T_{(0)} \right]= \end{equation}

\[=2Re\nabla _{z} W_{x} (-\frac{i}{4} )\left[\frac{Ra}{(1-i)^{2} } -1\right]\left[1-\frac{Ra}{(2i+Ra)^{*} } \right]\left|1-\frac{Ra}{(2i+Ra)} \right|^{2} A_{z} A_{z}^{*} .\]
Let us come now to the calculations of the term $2Re\nabla _{Z} \left\{\left[_{3} v_{(2)}^{z} \right]\left[_{4} T_{(0)} \right]\right\}:$

\begin{equation} \label{GEQ305} 2Re\nabla _{z} \left\{\overline{\left[_{3} v_{(2)}^{z} \right]\left[_{4} T_{(0)} \right]}\right\}= \end{equation}

\[=2Re\nabla _{z} \left\{-W_{q} [_{3} T_{(1)}^{qz} ][_{4} T_{(0)} ]-W_{q} [_{3} T_{(2)}^{qz} ][_{4} T_{(0)} ]\right\}.\]
Calculations similar to the previous ones give :

\begin{equation} \label{GEQ306} -W_{q} [_{3} T_{(1)}^{qz} ][_{4} T_{(0)} ]= \end{equation}

\[=\frac{(-i)RaW_{y} }{(1-i)^{4} (1+i)^{2} } \left[1-\frac{Ra}{(2i+Ra)^{*} } \right]\left|1-\frac{Ra}{(2i+Ra)} \right|^{2} B_{z} B_{z}^{*} ,\]

\begin{equation} \label{GEQ307} -W_{q} [_{3} T_{(2)}^{qz} ][_{4} T_{(0)} ]=\frac{iW_{y} }{4} \left[1-\frac{Ra}{(2i+Ra)^{*} } \right]\left|1-\frac{Ra}{(2i+Ra)^{*} } \right|^{2} B_{z} B_{z}^{*}  \end{equation}
As a result we obtain the term $\nabla _{z} \overline{v_{(2)}^{z} T_{(0)} }$:

\begin{equation} \label{GEQ308} \nabla _{z} \overline{v_{(2)}^{z} T_{(0)} }= \end{equation}

\[=2Re\nabla _{z} W_{x} (-\frac{i}{4} )\left[\frac{Ra}{(1-i)^{2} } -1\right]\left[1-\frac{Ra}{(2i+Ra)^{*} } \right]\left|1-\frac{Ra}{(2i+Ra)} \right|^{2} \left|A_{z} \right|^{2} +\]

\[+2Re\nabla _{z} W_{y} (-\frac{i}{4} )\left[\frac{Ra}{(1-i)^{2} } -1\right]\left[1-\frac{Ra}{(2i+Ra)^{*} } \right]\left|1-\frac{Ra}{(2i+Ra)^{*} } \right|^{2} \left|B_{z} \right|^{2} .\]
Let us come now to the calculations of the term $\overline{v_{(0)}^{z} T_{(2)} }$.

\begin{equation} \label{GEQ309} \nabla _{z} \overline{v_{(0)}^{z} T_{(2)} }=2Re\left\{\left[_{1} v_{(0)}^{z} \right]\left[_{2} T_{(2)} \right]+\left[_{3} v_{(0)}^{z} \right]\left[_{4} T_{(2)} \right]\right\}. \end{equation}
Let us write down the auxiliary expressions:
\begin{equation} \label{GEQ310} \left[_{2} T_{(2)} \right]=\frac{1}{\left(1+i\right)} \left(W_{q} [_{2} T_{(1)}^{qz} ]+W_{q} [_{2} T_{(2)}^{qz} ]+iW_{x} [_{2} T_{(0)} ]\right); \end{equation}

\begin{equation} \label{GEQ311} \left[_{4} T_{(2)} \right]\frac{1}{\left(1+i\right)} \left(W_{q} [_{4} T_{(1)}^{qz} ]+W_{q} [_{4} T_{(2)}^{qz} ]+iW_{y} [_{4} T_{(0)} ]\right). \end{equation}
Taking into account these formulae and expressions for $[_{2} T_{(1)}^{qz} ],[_{2} T_{(2)}^{qz} ],[_{4} T_{(1)}^{qz} ],[_{4} T_{(2)}^{qz} ],$ it is not difficult to get an expression for  the $\nabla _{z} \overline{v_{(0)}^{z} T_{(2)} }:$

\[\nabla _{z} \overline{v_{(0)}^{z} T_{(2)} }=\]

\[=-2Re\nabla _{z} \frac{iW_{x} }{2(1+i)^{2} } \left[-\frac{Ra}{(1+i)^{2} } \left(1-\frac{Ra}{(2i+Ra)} \right)+\left(1-\frac{Ra}{(2i+Ra)} \right)+1\right]\times \]

\[\times \left|1-\frac{Ra}{(2i+Ra)} \right|^{2} \left|A_{z} \right|^{2} -\]

\begin{equation} \label{GEQ312} -2Re\nabla _{z} \frac{iW_{y} }{2(1+i)^{2} } \times  \end{equation}

\[\times \left[-\frac{Ra}{(1+i)^{2} } \left(1-\frac{Ra}{(2i+Ra)} \right)+\left(1-\frac{Ra}{(2i+Ra)} \right)+1\right]\times \]

\[\times \left|1-\frac{Ra}{(2i+Ra)} \right|^{2} \left|B_{z} \right|^{2} .\]
The sum of the expressions (\ref{GEQ312}) and (\ref{GEQ308}) is the Reynolds stress (\ref{GEQ296}). After simple algebraical transformations  and finding of the real part we obtain the final expression:
\begin{equation} \label{GEQ313} \nabla _{Z} \left(\overline{v_{(2)}^{z} T_{(0)} }+\overline{v_{(0)}^{z} T_{(2)} }\right)=2\frac{Ra-2}{(4+Ra^{2} )^{2} } \nabla _{z} \left(W_{x} +W_{y} \right). \end{equation}

\section{Appendix D. Scheme of asymptotical development for the non linear case}

Let us present the algebraical structure of the asymptotical development of the equations  (\ref{GEQ119}), (\ref{GEQ120}) for the non linear theory (we will not write indices because they can be restored evidently at any moment). In the order $R^{-3} $ there is only the equation:

\begin{equation} \label{GEQ314} \partial P_{-3} =0,\Rightarrow P_{-3} =P_{-3} (X). \end{equation}
In the order $R^{-2} $ we have the equation :

\begin{equation} \label{GEQ315} \partial P_{-2} =0,\Rightarrow P_{-2} =P_{-2} (X). \end{equation}
In the order $R^{-1} $ we get a system of equations:

\begin{equation} \label{GEQ316} \partial _{t} W_{-1} -\partial ^{2} W_{-1} =-(\partial P_{-1} +\nabla P_{-3} )+RaT_{-1} l_{z} -\partial W_{-1} W_{-1} , \end{equation}

\begin{equation} \label{GEQ317} \partial _{t} T_{-1} -\partial ^{2} T_{-1} =-\partial W_{-1} T_{-1} -W_{-1}^{z} , \end{equation}

\[\partial W_{-1} =0.\]
The system of equations  (\ref{GEQ316}), (\ref{GEQ317}) gives secular terms:

\begin{equation} \label{GEQ318} -\nabla P_{-3} +RaT_{-1} l_{z} =0. \end{equation}

\begin{equation} \label{GEQ319} W_{-1}^{z} =0. \end{equation}
In zero order $R^{0} $ we have the following system of equations:

\begin{equation} \label{GEQ320} \partial _{t} v_{0} -\partial ^{2} v_{0} +\partial (W_{-1} v_{0} +v_{0} W_{-1} )= \end{equation}

\[=-(\partial P_{0} +\nabla P_{-2} )+RaT_{0} l_{z} +F,\]
\begin{equation} \label{GEQ321} \partial _{t} T_{0} -\partial ^{2} T_{0} +\partial (W_{-1} T_{0} +v_{0} T_{-1} )=-v_{0}^{z} . \end{equation}
\[\partial v_{0} =0.\]
These equations give one secular equation:

\begin{equation} \label{GEQ322} \nabla P_{-2} =0,\Rightarrow P_{-2} =Const. \end{equation}
Consider the equations of the first approximation $R$:

\begin{equation} \label{GEQ323} \partial _{t} v_{1} -\partial ^{2} v_{1} +\partial (W_{-1} v_{1} +v_{1} W_{-1} +v_{0} v_{0} )= \end{equation}

\[=-\nabla (W_{-1} W_{-1} )-(\partial P_{1} +\nabla P_{-1} )+RaT_{1} l_{z} .\]

\begin{equation} \label{GEQ324} \partial _{t} T_{1} -\partial ^{2} T_{1} +\partial (W_{-1} T_{1} +v_{1} T_{-1} +v_{0} T_{0} )+\nabla (W_{-1} T_{-1} )=-v_{1}^{z} . \end{equation}

\begin{equation} \label{GEQ325} \partial V_{1} +\nabla W_{-1} =0. \end{equation}
From this system of equations follow the secular equations:

\begin{equation} \label{GEQ326} \nabla W_{-1} =0, \end{equation}

\begin{equation} \label{GEQ327} \nabla (W_{-1} W_{-1} )=-\nabla P_{-1} , \end{equation}

\begin{equation} \label{GEQ328} \nabla (W_{-1} T_{-1} )=0. \end{equation}
The secular equations  (\ref{GEQ326})-(\ref{GEQ328}), are obviously satisfied in the previous velocity field geometry:

\begin{equation} \label{GEQ329} W=(W_{x} (Z),W_{y} (Z),0);T_{-1} =T_{-1} (Z);\nabla P_{-1} =0,\Rightarrow P_{-1} =Const. \end{equation}
In the second order $R^{2} $, we obtain equations :

\begin{equation} \label{GEQ330} \partial _{t} v_{2} -\partial ^{2} v_{2} -2\partial \nabla v_{0} +\partial (W_{-1} v_{2} +v_{2} W_{-1} +v_{0} v_{1} +v_{1} v_{0} )= \end{equation}

\[=-\nabla (W_{-1} v_{0} +v_{0} W_{-1} )-(\partial P_{2} +\nabla P_{0} )+RaT_{2} l_{z} ,\]

\begin{equation} \label{GEQ331} \partial _{t} T_{2} -\partial ^{2} T_{2} -2\partial \nabla T_{0} +\partial (W_{-1} T_{2} +v_{2} T_{-1} +v_{0} T_{1} +v_{1} T_{0} )= \end{equation}

\[=-\nabla (W_{-1} T_{0} +v_{0} T_{-1} )-v_{2} .\]

\begin{equation} \label{GEQ332} \partial v_{2} +\nabla v_{0} =0. \end{equation}
It is easy to see that in the order $R^{2} $ there is no secular terms.

Let us come  now to the most important order $R^{3} $. In this order we obtain equations:
\begin{equation} \label{GEQ333} \partial _{t} v_{3} +\partial _{T} W_{-1} -(\partial ^{2} v_{3} +2\partial \nabla v_{1} +\Delta W_{-1} )+\nabla (W_{-1} v_{1} +v_{1} W_{-1} +v_{0} v_{0} )+ \end{equation}

\[+\partial (W_{-1} v_{3} +v_{3} W_{-1} +v_{0} v_{2} +v_{2} v_{0} +v_{1} v_{1} )=-(\partial P_{3} +\nabla \overline{P}_{1} )+RaT_{3} l_{z} .\]
\begin{equation} \label{GEQ334} \partial _{t} T_{3} +\partial _{T} T_{-1} -(\partial ^{2} T_{3} +2\partial \nabla T_{1} +\Delta T_{-1} )+\nabla (W_{-1} T_{1} +v_{1} T_{-1} +v_{0} T_{0} )+ \end{equation}

\[+\partial (W_{-1} T_{3} +v_{3} T_{-1} +v_{0} T_{2} +v_{2} T_{0} +v_{1} T_{1} )=-v_{3}^{z} .\]

\[\partial v_{3} +\nabla v_{1} =0.\]
From this we get the main secular equation:

\begin{equation} \label{GEQ335} \partial _{T} W_{-1} -\Delta W_{-1} +\nabla \left(\overline{v_{0} v_{0} }\right)=-\nabla \overline{P}_{1} , \end{equation}

\begin{equation} \label{GEQ336} \partial _{T} T_{-1} -\Delta T_{-1} +\nabla (v_{0} T_{0} )=0. \end{equation}

\section{Appendix E. Reynolds stress in non linear case }

In order to calculate the Reynolds stresses we have first of all to calculate the expression:

\begin{equation} \label{GEQ337} \overline{v_{0}^{k} v_{0}^{i} }=2Re\left(\overline{v_{01}^{k} v_{01}^{i*} }+\overline{v_{03}^{k} v_{03}^{i*} }\right). \end{equation}
Taking into account the formula  (\ref{GEQ213}), we obtain:

\begin{equation} \label{GEQ338} \overline{v_{01}^{k} v_{01}^{i*} }+\overline{v_{01}^{k*} v_{01}^{i} }=T_{(1)}^{ki} =\frac{1}{\left|D_{1} \right|^{2} } (A_{k} A_{i}^{*} +A_{k}^{*} A_{i} )+ \end{equation}

\[-\frac{RaA_{z}^{*} }{\left|D_{1} \right|^{2} } \left(\frac{l_{k} A_{i} +l_{i} A_{k} }{D_{1}^{2} +Ra} \right)-\frac{RaA_{z} }{\left|D_{1} \right|^{2} } \left(\frac{l_{k} A_{i}^{*} +l_{i} A_{k}^{*} }{D_{1}^{*2} +Ra} \right)+\]

\[+\frac{2}{\left|D_{1} \right|^{2} } \frac{Ra^{2} l_{k} l_{i} \left|A_{z} \right|^{2} }{\left|D_{1}^{2} +Ra\right|^{2} } .\]
Similarly taking into account the formula  (\ref{GEQ214}), we obtain:

\begin{equation} \label{GEQ339} \overline{v_{03}^{k} v_{03}^{i*} }+\overline{v_{03}^{k*} v_{03}^{i} }=T_{(2)}^{ki} =\frac{1}{\left|D_{2} \right|^{2} } (B_{k} B_{i}^{*} +B_{k}^{*} B_{i} )- \end{equation}

\[-\frac{RaB_{z}^{*} }{\left|D_{2} \right|^{2} } \left(\frac{l_{k} B_{i} +l_{i} B_{k} }{D_{2}^{2} +Ra} \right)-\frac{RaB_{z} }{\left|D_{2} \right|^{2} } \left(\frac{l_{k} B_{i}^{*} +l_{i} B_{k}^{*} }{D_{2}^{*2} +Ra} \right)+\]

\[+\frac{2}{\left|D_{2} \right|^{2} } \frac{Ra^{2} l_{k} l_{i} \left|B_{z} \right|^{2} }{\left|D_{2}^{2} +Ra\right|^{2} } .\]
It is clear that the components $T_{(1)}^{3i} $ and $T_{(2)}^{3i} $ are of interest.  To begin with we consider the components  of the tensor $T_{(1)}^{3i} $.

\begin{equation} \label{GEQ340} T_{(1)}^{31} =\frac{1}{\left|D_{1} \right|^{2} } (A_{3} A_{1}^{*} +A_{3}^{*} A_{1} )- \end{equation}

\[-\frac{Ra}{\left|D_{1} \right|^{2} } \left(\frac{A_{3}^{*} A_{1} }{D_{1}^{2} +Ra} +\frac{A_{3} A_{1}^{*} }{D_{1}^{*2} +Ra} \right)=0,\]
Since$A_{3} A_{1}^{*} =A_{3}^{*} A_{1} =0.$
\begin{equation} \label{GEQ341} T_{(1)}^{32} =\frac{1}{\left|D_{1} \right|^{2} } (A_{3} A_{2}^{*} +A_{3}^{*} A_{2} )- \end{equation}

\[-\frac{Ra}{\left|D_{1} \right|^{2} } \left(\frac{A_{3}^{*} A_{2} }{D_{1}^{2} +Ra} +\frac{A_{3} A_{2}^{*} }{D_{1}^{*2} +Ra} \right).\]
The first bracket in the (\ref{GEQ341}) is equal to zero, which is why:

\begin{equation} \label{GEQ342} T_{(1)}^{32} =-\frac{i}{4} \frac{Ra}{\left|D_{1} \right|^{2} } \frac{\left(D_{1}^{2} -D_{1}^{*2} \right)}{\left|D_{1}^{2} +Ra\right|^{2} } . \end{equation}
Now consider the component $T_{(2)}^{32} :$

\begin{equation} \label{GEQ343} T_{(2)}^{32} =\frac{1}{\left|D_{2} \right|^{2} } (B_{3} B_{2}^{*} +B_{3}^{*} B_{2} )- \end{equation}

\[-\frac{Ra}{\left|D_{2} \right|^{2} } \left(\frac{B_{3}^{*} B_{2} }{D_{1}^{2} +Ra} +\frac{B_{3} B_{2}^{*} }{D_{1}^{*2} +Ra} \right)=0.\]
As far as  $B_{3}^{*} B_{2} =B_{3} B_{2}^{*} =0$ we consider the component $T_{(2)}^{31} $:

\begin{equation} \label{GEQ344} T_{(2)}^{31} =\frac{1}{\left|D_{2} \right|^{2} } (B_{3} B_{1}^{*} +B_{3}^{*} B_{1} )- \end{equation}

\[-\frac{Ra}{\left|D_{2} \right|^{2} } \left(\frac{B_{3}^{*} B_{1} }{D_{2}^{2} +Ra} +\frac{B_{3} B_{1}^{*} }{D_{2}^{*2} +Ra} \right).\]
The first bracket in the formula(\ref{GEQ344}) is equal to zero, then:

\begin{equation} \label{GEQ345} T_{(2)}^{31} =-\frac{i}{4} \frac{Ra}{\left|D_{2} \right|^{2} } \frac{\left(D_{2}^{*2} -D_{2}^{2} \right)}{\left|D_{2}^{2} +Ra\right|^{2} } . \end{equation}
Taking into account :

\begin{equation} \label{GEQ346} \left(D_{1}^{2} -D_{1}^{*2} \right)=4i(1-W_{1} ),\left(D_{2}^{*2} -D_{2}^{2} \right)=-4i(1-W_{2} ), \end{equation}

\begin{equation} \label{GEQ347} \left|D_{1} \right|^{2} =1+(1-W_{1} );\left|D_{2} \right|^{2} =1+(1-W_{2} ), \end{equation}

\begin{equation} \label{GEQ348} \left|D_{1}^{2} +Ra\right|^{2} =\left(W_{1} (2-W_{1} )+Ra\right)^{2} +4(1-W_{1} )^{2} , \end{equation}
\begin{equation} \label{GEQ349} \left|D_{2}^{2} +Ra\right|^{2} =\left(W_{2} (2-W_{2} )+Ra\right)^{2} +4(1-W_{2} )^{2} . \end{equation}
The components $T_{(1)}^{32} $, $T_{(2)}^{31} $ take the form:

\begin{equation} \label{GEQ350} T_{(1)}^{32} =\frac{Ra(1-W_{1} )}{[1+(1-W_{1} )^{2} ][\left(W_{1} (2-W_{1} )+Ra\right)^{2} +4(1-W_{1} )^{2} ]} , \end{equation}

\begin{equation} \label{GEQ351} T_{(2)}^{31} =-\frac{Ra(1-W_{2} )}{[1+(1-W_{2} )^{2} ][\left(W_{2} (2-W_{2} )+Ra\right)^{2} +4(1-W_{2} )^{2} ]} . \end{equation}
Now calculate the components $T_{(1)}^{33} $ and $T_{(2)}^{33} $. It is easy to see that:

\begin{equation} \label{GEQ352} T_{(1)}^{33} =\frac{2\left|A_{3} \right|^{2} }{\left|D_{1} \right|^{2} } \left[1-Ra\frac{\left(D_{1}^{2} +D_{1}^{*2} \right)}{\left|D_{1}^{2} +Ra\right|^{2} } -\frac{Ra^{2} }{\left|D_{1}^{2} +Ra\right|^{2} } \right], \end{equation}

\begin{equation} \label{GEQ353} T_{(2)}^{33} =\frac{2\left|B_{3} \right|^{2} }{\left|D_{2} \right|^{2} } \left[1-Ra\frac{\left(D_{2}^{2} +D_{2}^{*2} \right)}{\left|D_{2}^{2} +Ra\right|^{2} } -\frac{Ra^{2} }{\left|D_{2}^{2} +Ra\right|^{2} } \right]. \end{equation}
Or in the explicit form:

\begin{equation} \label{GEQ354} T_{(1)}^{33} =\frac{1}{2[1+(1-W_{1} )^{2} ]} \times  \end{equation}

\[\times \left[1-Ra\frac{2[1-(1-W_{1} )^{2} ]+Ra}{[W_{1} (2-W_{1} )+Ra]^{2} +4(1-W_{1} )^{2} } \right],\]

\begin{equation} \label{GEQ355} T_{(2)}^{33} =\frac{1}{2[1+(1-W_{2} )^{2} ]} \times  \end{equation}

\[\times \left[1-Ra\frac{2[1-(1-W_{2} )^{2} ]+Ra}{[W_{2} (2-W_{2} )+Ra]^{2} +4(1-W_{2} )^{2} } \right].\]

\end{document}